\documentclass[aps,prx,reprint,nofootinbib,floatfix,longbibliography]{revtex4-2}

\usepackage{amssymb}
\usepackage{amsmath}
\usepackage{amsfonts}
\usepackage{graphicx}
\usepackage{dcolumn}
\usepackage{bm}
\usepackage{color}
\usepackage{booktabs}
\usepackage{multirow}

\begin{document}

\title{Where Is the \(p\)-Wave Superfluid? Correlated Phase Diagram of Microwave-Shielded Polar Molecules}

\author{Tao Shi}
\affiliation{Institute of Theoretical Physics, Chinese Academy of Sciences,
P.O. Box 2735, Beijing 100190, China}

\date{\today}

\begin{abstract}
Chiral $p_x\!\pm\! i p_y$ superfluidity in fermionic polar molecules offers a direct route to topological quantum matter with non-Abelian excitations, yet the parameter regime in which pairing survives strong correlations and competing instabilities has remained unresolved. Here we determine this regime for microwave-shielded polar molecules. Under tight confinement along the \(z\) direction, the microscopic molecular interaction collapses onto a universal quasi-two-dimensional form specified by an interaction length $\sigma$, defined as the position of its attractive minimum, and a single dimensionless coupling. We combine Fermi-hypernetted-chain Euler--Lagrange theory with correlated-basis pairing theory, thereby treating the correlation hole, density response, stability of the homogeneous fluid, and $p$-wave pairing within one framework. The largest pairing gap occurs immediately on the stable side of a long-wavelength spinodal instability at a two-dimensional density $4\times10^{-2}/\sigma^2$. Mapping this universal regime onto microscopic shielding schemes exposes an intrinsic limitation of single-microwave dressing: weakening the attraction to the optimal pairing range simultaneously weakens collisional shielding. Dual-microwave dressing removes this constraint by compensating the long-range attraction while retaining a large repulsive core. Imposing explicit requirements on collisional loss, confinement, and microwave-amplitude stability, we identify the regime of molecular mass and dipole moment favorable for \(p\)-wave superfluidity and determine experimentally realistic microwave and confinement conditions for its realization. These results turn the search for molecular $p$-wave superfluidity into a quantitative optimization problem and define a route toward the topological weak-pairing phase.
\end{abstract}

\maketitle

\section{Introduction}
\label{sec:introduction}

Ultracold polar molecules combine long-lived internal states with strong, anisotropic, and externally tunable dipole--dipole interactions, providing a platform for precision measurement, quantum simulation, controlled chemistry, and quantum information science~\cite{Bohn2017,DeMille2017,Langen2024,Cornish2024,Schindewolf2026RMP}. For itinerant fermionic molecules, a particularly long-standing objective is the chiral $p_x\!\pm\! i p_y$ superfluid. Its weak-pairing phase is topological and supports vortex-bound Majorana modes with non-Abelian exchange statistics, directly connecting a mobile molecular gas to the physics sought for topological quantum information processing~\cite{ReadGreen2000,Ivanov2001,Nayak2008}. The broader pursuit of odd-parity pairing in dipolar Fermi gases spans more than two decades~\cite{Baranov2002,Baranov2004,Shi2010,Shi2013}, while microwave-dressed polar molecules offer an unusually direct way to engineer the attractive interaction required for a topological $p$-wave phase~\cite{CooperShlyapnikov2009,Levinsen2011,Buechler2007}. Microwave shielding has now made this route experimentally concrete by suppressing short-range collisional loss while retaining strong, tunable interactions~\cite{Gorshkov2008,KarmanHutson2018,KarmanHutson2019,Anderegg2021}. Fermionic $^{23}\mathrm{Na}^{40}\mathrm{K}$ molecules have been evaporatively cooled into the quantum-degenerate regime~\cite{Schindewolf2022}; field-linked resonances have enabled coherent association into ultracold tetramers~\cite{Chen2023Field,Chen2024Tetramer}; and, most recently, double-microwave shielding produced a deeply degenerate Fermi gas in which interaction-induced Fermi-surface deformation was directly observed~\cite{Biswas2026FermiSurface}. Important advances have also been achieved with bosonic polar molecules. Microwave shielding stabilized dense NaCs and NaRb gases~\cite{Bigagli2023,Lin2023}, double-microwave dressing enabled Bose--Einstein condensation of NaCs~\cite{Bigagli2024}, and subsequent experiments demonstrated broad interaction tunability, extreme loss suppression, a tunable NaRb condensate, and self-bound molecular droplets~\cite{Yuan2025Loss,Shi2026NaRb,Zhang2026Droplet}. Microwave-shielded polar molecules (MSPMs) have therefore entered a regime in which the search for fermionic topological superfluidity can be posed as a quantitative many-body and experimental-design problem.

These experimental advances have been accompanied by an increasingly mature
many-body description on the bosonic side, whereas the corresponding theory for
fermionic microwave-shielded molecules remains far less developed. For a single
circularly polarized microwave, the effective molecular interaction contains a
tunable attractive dipolar tail together with a large repulsive shielding
core~\cite{Deng2023}; related effective descriptions for dual-microwave dressing
have since been established and benchmarked by coupled-channel and universality
analyses~\cite{Deng2025Dual,Karman2025Double,Dutta2025,XuChen2025}. These
interactions now underpin much of the theoretical description of
microwave-shielded gases, as summarized in the recent review of
Ref.~\cite{Schindewolf2026RMP}. Importantly, the shielding core is not a minor
short-range correction. Once its size becomes comparable to the mean
intermolecular spacing, many-body correlations become substantial and a
description based solely on low-energy scattering parameters is no longer
sufficient. For bosons, this observation has led to a liquid-$^4$He-like
viewpoint and to a sequence of correlated studies beyond the Gross--Pitaevskii description, including variational and quantum Monte Carlo
calculations of molecular droplets, finite-temperature fluids, supersolids, and
crystalline phases~\cite{Jin2025,Langen2025,Ciardi2025,Polterauer2025,
Zhang2025FiniteT,SanchezBaena2025,Zhang2025Supersolid,Cardinale2026,
Ciardi2026Nonequilibrium}. Recent few-body work has further shown that strongly
anisotropic microwave dressing can support universal molecular bound states
with an emergent Bose--Fermi duality~\cite{ShiWangCui2026}. For fermions, by contrast, previous studies of superfluidity have remained
largely within weak-coupling Bardeen–Cooper–Schrieffer (BCS) theory~\cite{Deng2023}. In the strongly
shielded regime relevant to current experiments, however, the problem is much
closer in spirit to liquid $^3$He: the correlation hole substantially
renormalizes the quasiparticle interaction, while increasing the attraction simultaneously enhances long-wavelength density fluctuations and eventually drives the homogeneous fluid toward a spinodal instability associated with droplet formation. Pairing therefore
cannot be optimized independently of the correlated normal state and its
stability. A complete many-body theory that treats strong correlations,
$p$-wave pairing, and the competing density instability on equal footing is
therefore needed to determine where a stable molecular $p$-wave superfluid can
actually exist.

Here we develop a correlated many-body description based on the
Fermi-hypernetted-chain Euler--Lagrange (FHNC-EL) and correlated-basis
framework~\cite{Krotscheck1977,Krotscheck2000,Fan2015,FanKrotscheck2019}
to determine where a stable molecular $p$-wave superfluid can be realized
in a quasi-two-dimensional (quasi-2D) geometry. This framework has been
extensively developed for strongly correlated bulk liquid $^3$He, where
short-range correlations and fermionic pairing must likewise be treated
beyond conventional mean-field theory. We first show that, under sufficiently
tight confinement along the $z$ direction, the quasi-2D microwave-shielded
interaction becomes universal: after rescaling lengths by the position
$\sigma$ of its attractive minimum, the interaction is characterized by a
single dimensionless coupling $\bar{C}_0$. We then optimize the correlated
normal state within FHNC-EL theory and solve the pairing problem in the
resulting correlated basis. The phase diagram reveals a simple but nontrivial principle: the pairing gap
is maximized not at the strongest available attraction, but immediately on
the stable side of a long-wavelength spinodal instability associated with
droplet formation. The maximal zero-temperature gap,
$\Delta_{\max}/\varepsilon_F\simeq0.036$, occurs near
$k_F\sigma\simeq0.7$ and $\bar{C}_0\simeq3$--$4$, where $k_F$ and
$\varepsilon_F$ denote the Fermi momentum and Fermi energy, respectively.
We next map this universal window back to realistic molecular interactions.
For single-microwave shielding, the attraction and collisional protection
originate from the same dressing field, making it difficult to retain strong
shielding while reducing the attraction to the moderate strength required by
the correlated phase diagram. Dual-microwave shielding separates these two
requirements. By balancing pairing strength, stability against droplet formation,
collisional loss, confinement, and sensitivity to
microwave-amplitude fluctuations, we identify the range of molecular masses
and dipole moments that favors a stable $p$-wave superfluid and determine
experimentally accessible microwave and confinement conditions for its realization.

The paper is organized as follows. Section~II develops the FHNC-EL and
correlated-basis formulation. Section~III introduces the universal quasi-2D interaction and
establishes the quasi-2D phase diagram. Section~IV connects this phase diagram to realistic single- and
dual-microwave shielding schemes and identifies experimentally accessible
working regimes. Section~V summarizes the resulting design principles and discusses the
finite-temperature requirements for realizing the $p$-wave superfluid.

\section{FHNC and correlated-basis theory of pairing}
We determine superfluid pairing on top of an optimized correlated normal state. The FHNC-EL calculation treats the short-distance correlation hole generated by the shielding core and the long-wavelength density response on the same footing. Pairing is then formulated in the associated correlated basis, so that the bare molecular potential is replaced by a medium- and correlation-renormalized interaction. In the weak BCS pairing regime relevant below, feedback of the gap on the optimized normal-state correlations enters only at second order in the gap and is neglected. Throughout this section we set $\hbar=1$ in the FHNC equations and restore it explicitly when quoting physical energy scales and dimensionless couplings.

For a $\nu$-component molecular gas of mass $M$ in $d$ dimensions, the
Hamiltonian reads
\begin{equation}
\begin{aligned}
H={}&
-\frac{1}{2M}
\sum_{\sigma}
\int d^d r\,
\psi_\sigma^\dagger(\mathbf r)
\nabla^2
\psi_\sigma(\mathbf r)
\nonumber\\
&\quad
+\frac{1}{2}
\int d^d r_1 d^d r_2\,
V(\mathbf r_1-\mathbf r_2)\,
\mathopen{:}
n(\mathbf r_1)n(\mathbf r_2)
\mathclose{:},
\label{eq:H}
\end{aligned}
\end{equation}
where $V(\mathbf r)$ is the two-body molecular interaction, $n(\mathbf{r}%
)=\sum_{\sigma }\psi _{\sigma }^{\dagger }(\mathbf{r})\psi _{\sigma }(%
\mathbf{r})$ is the density operator, and $\mathopen{:}\cdots\mathclose{:}$ denotes
normal ordering. To describe the correlated normal state, we use the
Jastrow-Feenberg variational ansatz
$|\Psi _{N}\rangle =e^{\mathcal{S}}|%
\mathrm{FS}\rangle /\sqrt{I_{0}}$,
where $|{\rm FS}\rangle$ denotes the noninteracting Fermi sea with Fermi
momentum $k_F$, and
$I_0=\langle {\rm FS}|e^{2\mathcal S}|{\rm FS}\rangle$
is the normalization factor. The correlation operator is taken to be
\begin{equation}
    \mathcal S =
    \frac{1}{4}
    \int d^d r_1 d^d r_2\,
    u(\mathbf r_1-\mathbf r_2)\,
    \mathopen{:}
    n(\mathbf r_1)n(\mathbf r_2)
    \mathclose{:},
    \label{eq:JF_generator}
\end{equation}
with $u(\mathbf r)$ the two-body correlation function optimized by the
FHNC-EL equations~\cite{Krotscheck1977,Krotscheck2000,Fan2015,FanKrotscheck2019}.

\subsection{Optimized correlated normal state and stability}
The pair distribution function is defined as
$g(\mathbf r)=
\langle\Psi_N|\mathopen{:}n(\mathbf r)n(\mathbf 0)\mathclose{:}|\Psi_N\rangle/
\rho^2$, where $\rho$ is the total density. Minimizing the normal-state
energy $E_N=\langle\Psi_N|H|\Psi_N\rangle$ with respect to the correlation
function gives the EL relation
\begin{equation}
    g'(\mathbf r)=\frac{1}{4M}\nabla^2 g(\mathbf r).
    \label{gprime}
\end{equation}
The prime is not a derivative with respect to the function argument. It denotes the kinetic-energy numerator associated with the corresponding FHNC diagram, obtained through the Jackson--Feenberg (JF) identity~\cite{Krotscheck1977,Krotscheck2000,FanKrotscheck2019}. Fourier transforming
Eq.~\eqref{gprime} gives the JF derivative
\begin{equation}
    S'(\mathbf k)=-\frac{k^2}{4M}\left[S(\mathbf k)-1\right]
    \label{Sprime}
\end{equation}
of the static structure factor
\[
S(\mathbf k)=1+\rho\int d^d r\,e^{-i\mathbf k\cdot\mathbf r}
[g(\mathbf r)-1].
\]
Here, we use the density-normalized
Fourier convention
$f(\mathbf k)=\rho\int d^d r\,e^{-i\mathbf k\cdot\mathbf r}f(\mathbf r)$,
with inverse
$f(\mathbf r)=\rho^{-1}\int d^d k\,e^{i\mathbf k\cdot\mathbf r}
f(\mathbf k)/(2\pi)^d$. For compactness, the same symbol is used for a
function in coordinate and momentum space, and its argument specifies the
representation.

Within FHNC theory, the density correlation function can be factorized as
$g(\mathbf r)=g_{\rm dd}(\mathbf r)g_{\rm s}(\mathbf r)$, where
$g_{\rm dd}(\mathbf r)=1+\Gamma_{\rm dd}(\mathbf r)$ is the dynamical
correlation generated by the Jastrow-Feenberg correlation operator, while
$g_{\rm s}(\mathbf r)=g_F(\mathbf r)+C(\mathbf r)$ contains the statistical
correlations imposed by Fermi exchange. The noninteracting contribution is
$g_F(\mathbf r)=1-l_0^2(\mathbf r)/\nu$, with the normalized single-particle
correlation function
$l_0(\mathbf r)=\nu\langle{\rm FS}|
\psi_\sigma^\dagger(\mathbf 0)\psi_\sigma(\mathbf r)
|{\rm FS}\rangle/\rho$:
\begin{equation}
    l_0(\mathbf r)=
    \frac{\nu}{\rho}
    \int_{|\mathbf k|<k_F}
    \frac{d^d k}{(2\pi)^d}
    e^{i\mathbf k\cdot\mathbf r}.
\end{equation}
For $d=2$ and $d=3$,
$l_0(r)=2J_1(k_Fr)/(k_Fr)$ and
$l_0(r)=3j_1(k_Fr)/(k_Fr)$, respectively, where
$r=|\mathbf r|$. The exchange correction $C(\mathbf r)$ accounts for the
leading FHNC exchange diagrams beyond the ideal-gas statistical factor. Using the FHNC cluster expansion, the momentum-space correlations are related
to the static structure factor by
\begin{equation}
\begin{aligned}
\Gamma_{\rm dd}(\mathbf k)
&=
\frac{S(\mathbf k)-S_\sigma(\mathbf k)}
     {S_\sigma^2(\mathbf k)},
\\
C(\mathbf k)
&=
\Gamma_{\rm dd}(\mathbf k)
\left[S_\sigma^2(\mathbf k)-1\right]
+\delta X^{(3,4)}_{\rm ee}(\mathbf k),
\label{GFHNC}
\end{aligned}
\end{equation}
where the spin structure function $S_\sigma(\mathbf k)=S_F(\mathbf k)+\delta X_{\rm ee}(\mathbf k)$ is composed of the static structure factor $S_F(\mathbf k)=1-\rho\int d^d r\,e^{-i\mathbf k\cdot\mathbf r}
l_0^2(\mathbf r)/\nu$ for the noninteracting
Fermi gas (see Appendix~\ref{app:SFI} for analytical forms) and the leading exchange correction $\delta X_{\rm ee}(\mathbf k)=\delta X_{\mathrm{ee}}^{(2)}(\mathbf{k})+\delta X_{\mathrm{ee}}^{(3,4)}(\mathbf{k})$. In \(\delta X_{\rm ee}(\mathbf{k})\), we retain the two- through
four-body contributions
\begin{equation}
\begin{aligned}
\delta X_{\rm ee}^{(2)}(\mathbf{k})
&=
-\frac{\rho}{\nu}
\int d\mathbf{r}\,
e^{-i\mathbf{k}\cdot\mathbf{r}}
\Gamma_{\rm dd}(\mathbf{r})l_0^2(\mathbf{r}),
\\
\delta X_{\rm ee}^{(3,4)}(\mathbf{k})
&=-\frac{\rho}{\nu}
\int d\mathbf{r}\,
\Gamma_{\rm dd}(\mathbf{r})[|I(\mathbf{k},
\mathbf{r})|^2\\
&\quad
-2l_0(\mathbf{r})
\operatorname{Re}I(\mathbf{k},\mathbf{r})],
\label{Xee}
\end{aligned}
\end{equation}
where the explicit evaluation of the exchange convolution
\begin{equation}
    I(\mathbf k,\mathbf r)=
    \frac{\rho}{\nu}
    \int d^d r'\,
    e^{-i\mathbf k\cdot\mathbf r'}
    l_0(\mathbf r')l_0(\mathbf r-\mathbf r'),
\end{equation}
is summarized in Appendix~\ref{app:SFI}. The exchange contribution $\delta X_{\mathrm{ee}}(\mathbf{k})$ is essential rather than optional: the cancellations imposed by Fermi statistics control the long-wavelength limit and are required for a stable FHNC-EL optimization~\cite{Krotscheck1977,Krotscheck2000,FanKrotscheck2019}.

Combining the EL relation in Eq.~\eqref{Sprime}, the JF prime
$S_F'(\mathbf k)=-k^2[S_F(\mathbf k)-1]/(4M)$ of the ideal-gas structure factor, and the FHNC relation for
$\Gamma_{\rm dd}(\mathbf k)$ in Eq.~\eqref{GFHNC}, one obtains the
FHNC-EL equation~\cite{FanKrotscheck2019}
\begin{equation}
S(\mathbf k)=
\frac{S_\sigma(\mathbf k)}
{\sqrt{D(\mathbf k)}},
\qquad
D(\mathbf k)=1+\frac{4M S_\sigma^2(\mathbf k)}{k^2}
\bar V(\mathbf k),
\label{eq:FHNC_EL_main}
\end{equation}
where $\bar V(\mathbf k)=V_{\rm ph}(\mathbf k)+V_{\rm ex}(\mathbf k)$. The particle-hole interaction is most
conveniently written in coordinate space as
\begin{equation}
V_{\mathrm{ph}}(\mathbf{r})=\frac{1}{M}\left\vert \nabla \sqrt{g_{\mathrm{dd}%
}(\mathbf{r})}\right\vert ^{2}+V(\mathbf{r})g_{\mathrm{dd}}(\mathbf{r})+w_{%
\mathrm{I}}(\mathbf{r})\Gamma _{\mathrm{dd}}(\mathbf{r}),
\label{Vph}
\end{equation}%
where the induced interaction in the momentum space is
\begin{equation}
\begin{aligned}
w_{\rm I}(\mathbf k)
={}&
-\frac{k^2}{4M}
\left[
\frac{1}{S_\sigma(\mathbf k)}
-\frac{1}{S(\mathbf k)}
\right]^2
\left[
1+2\frac{S(\mathbf k)}{S_\sigma(\mathbf k)}
\right]
\\
&\quad
-\frac{2}{S_\sigma(\mathbf k)}
V_{\rm ee}(\mathbf k)\Gamma_{\rm dd}(\mathbf k).
\end{aligned}
\end{equation}
The exchange contribution $V_{\rm ex}(\mathbf k)=V_{\rm ee}(\mathbf k)/S_\sigma^2(\mathbf k)$
is generated by the JF prime of
$\delta X_{\rm ee}(\mathbf k)$. As a result, the quantity $V_{\rm ee}(\mathbf k)$ has the same exchange structure as
$\delta X_{\rm ee}(\mathbf k)$ in Eq.~\eqref{Xee}, with
$\Gamma_{\rm dd}(\mathbf r)$ replaced by an effective interaction
$W(\mathbf r)$. Its Fourier transform is
\begin{equation}
W(\mathbf k)=
-\frac{k^2\Gamma_{\rm dd}(\mathbf k)}
{2M S_\sigma(\mathbf k)}
+
\left[
1-2\frac{S(\mathbf k)}{S_\sigma(\mathbf k)}
\right]
V_{\rm ex}(\mathbf k).
\label{Wk}
\end{equation}
Since $W(\mathbf k)$ itself contains $V_{\rm ex}(\mathbf k)$, the exchange
potential $V_{\rm ex}(\mathbf k)$ is, in principle, determined self-consistently. In the present
calculation we keep the leading contribution by evaluating $V_{\rm ee}$ with $W(\mathbf k)\sim W_0(\mathbf k)=
-k^2\Gamma_{\rm dd}(\mathbf k)/[2M S_\sigma(\mathbf k)]$.

Equation~\eqref{eq:FHNC_EL_main} is solved iteratively and provides the central stability criterion for the correlated normal state. In the long-wavelength limit, $\delta X_{\rm ee}(\mathbf k)
\sim k^2$, $S_F(\mathbf k)\sim c_F k/k_F$, and the optimized structure factor
takes the phononic form $S(\mathbf k)\to \hbar k/(2Mc)$. Within the collective approximation the density
excitation spectrum is $\omega _{\mathbf{k}}=\hbar k^{2}/(2MS)=ck$, thus, the
coefficient $c=v_{F}\sqrt{1+2c_{F}^{2}\bar{V}(0)/\varepsilon _{F}}%
/(2c_{F})$ is the sound velocity, where $v_{F}=\hbar k_F/M$ is the Fermi velocity and $\varepsilon
_{F}=\hbar ^{2}k_{F}^{2}/2M$ is the Fermi energy. A physical homogeneous FHNC-EL solution requires the argument of the square root in Eq.~\eqref{eq:FHNC_EL_main} to remain positive for all momenta. Its vanishing at \(k=0\), equivalently \(c=0\), marks a long-wavelength spinodal instability associated with collapse of the homogeneous state and possible droplet formation, whereas a zero at finite momentum signals an instability toward a spatially modulated density state. In addition, upon entering the strong-coupling regime on the BEC side, the in-medium particle-particle scattering amplitude can develop a pole,
signaling the formation of molecular dimers and the breakdown of the
correlated Fermi-liquid reference state. The present work focuses on the
BCS side, where no such bound-state instability occurs. Pairing calculations
are therefore performed within the parameter regime in which the optimized
normal-state solution remains regular and stable against density instabilities.

In the correlated normal state, the
dispersion of a fermion dressed by its correlation hole is
\begin{equation}
\varepsilon_{\mathbf k}^{(0)}
=
\frac{k^2}{2M}
-\mu_0
+
\frac{
X_{cc}^{(0)\prime}(\mathbf k)
}{
1-X_{cc}^{(0)}(\mathbf k)
}.
\end{equation}
The correlation-induced self-energy is determined by
\begin{equation}
\begin{aligned}
X_{cc}^{(0)}(\mathbf k)
&=
-\int
\frac{d^d p}{(2\pi)^d}\,
\Gamma_{\rm dd}(\mathbf k-\mathbf p)
l_0(\mathbf p),
\\
X_{cc}^{(0)\prime}(\mathbf k)
&=
-\int
\frac{d^d p}{(2\pi)^d}\,
W(\mathbf k-\mathbf p)
l_0(\mathbf p).
\label{Xcc}
\end{aligned}
\end{equation}
Here, the chemical potential $\mu_0$ is fixed by the density constraint
$\int d^d k\,\theta[-\varepsilon_{\mathbf k}^{(0)}]/(2\pi)^d=\rho/\nu$. For functions $\Gamma_{\rm dd}(\mathbf k-\mathbf p)$ and $W(\mathbf k-\mathbf p)$ depending on two momenta, we use the standard Fourier transform as $f(\mathbf{k}-\mathbf{p})=\int d^{d}r\,
e^{-i(\mathbf{k}-\mathbf{p})\cdot\mathbf{r}}f(r)$.
This optimized state, including its dressed dispersion and correlation-renormalized interactions, is the reference state for the pairing calculation.

\subsection{Pairing in the correlated basis}
To describe fermionic superfluidity in the presence of strong correlations,
we use the correlated BCS construction
\begin{equation}
|\Psi_{\rm CBCS}\rangle
=
\sum_{N,n}
\langle n^{(N)}|{\rm BCS}\rangle\,
\frac{e^{\mathcal S}|n^{(N)}\rangle}{\sqrt{I_n}} .
\label{eq:CBCS_appendix}
\end{equation}
Here, the uncorrelated BCS state $|{\rm BCS}\rangle$ is expanded in
fixed-particle-number Slater determinants $|n^{(N)}\rangle$. The same correlation operator $\mathcal S$ as in Eq.~\eqref{eq:JF_generator} is then
applied to each configuration, with
$I_n=\langle n^{(N)}|e^{2\mathcal S}|n^{(N)}\rangle$ the corresponding
normalization factor.

The normal and anomalous single-particle correlation functions of the BCS
state are defined as
$l_v(\mathbf r)=\nu\langle
\psi^\dagger(\mathbf 0)\psi(\mathbf r)\rangle_{\rm BCS}/\rho$ and
$l_u(\mathbf r)=\nu\langle
\psi(\mathbf r)\psi(\mathbf 0)\rangle_{\rm BCS}/\rho$, respectively. Their
Fourier representations are
\begin{equation}
\begin{aligned}
l_v(\mathbf r)
&=
\frac{\nu}{\rho}
\int \frac{d^d k}{(2\pi)^d}\,
l_v(\mathbf k)e^{i\mathbf k\cdot\mathbf r},
\\
l_u(\mathbf r)
&=
\frac{\nu}{\rho}
\int \frac{d^d k}{(2\pi)^d}\,
l_u(\mathbf k)e^{i\mathbf k\cdot\mathbf r}.
\end{aligned}
\label{eq:lvu_appendix}
\end{equation}
where $l_v(\mathbf k)=|v_{\mathbf k}|^2$ and
$l_u(\mathbf k)=u_{\mathbf k}v_{\mathbf k}^\ast$ are determined by the
Bogoliubov amplitudes $u_{\mathbf k}$ and $v_{\mathbf k}$. In the weak-pairing regime, the superfluid gap $|\Delta|\ll\varepsilon_F$, the feedback of the pairing gap on $S_\sigma$, $\Gamma_{\rm dd}$, and $W$ is of order $(|\Delta|/\varepsilon_F)^2$ and is neglected~\cite{Fan2015,FanKrotscheck2019}.

The FHNC cluster expansion yields the ground-state energy of $|\Psi_{\rm CBCS}\rangle$. Functional differentiation with respect to
$l_v(\mathbf k)$ and $l_u(\mathbf k)$ gives the renormalized dispersion
\begin{equation}
\mathcal E_{\mathbf k}
=
\frac{k^2}{2M}
-\mu
+
\frac{X_{\rm cc}'(\mathbf k)}
     {1-X_{\rm cc}(\mathbf k)}
+
\mathcal E_{\mathbf k}^{(u)},
\end{equation}
and the pairing gap
\begin{equation}
\Delta_{\mathbf k}
=
-\int
\frac{d^d p}{(2\pi)^d}\,
V_{\rm pp}(\mathbf k,\mathbf p)
l_u(\mathbf p).
\label{Dk}
\end{equation}
Here, $X_{\rm cc}(\mathbf k)$ and $X_{\rm cc}'(\mathbf k)$ have the same form
as in Eq.~\eqref{Xcc}, with $l_0(\mathbf p)$ replaced by $l_v(\mathbf p)$.
The additional contribution generated by the anomalous correlations is
\[
\mathcal E_{\mathbf k}^{(u)}
=
-2\,{\rm Re}\left[
\varepsilon_{\mathbf k}^{(0)}
l_u^\ast(\mathbf k)
\int
\frac{d^d p}{(2\pi)^d}\,
\Gamma_{\rm dd}(\mathbf k-\mathbf p)
l_u(\mathbf p)
\right].
\]

The correlated-basis effective interaction responsible for pairing is $V_{\rm pp}(\mathbf k,\mathbf p)=W(\mathbf k-\mathbf p)+V_\Gamma(\mathbf k,\mathbf p)$, where
\begin{equation}
V_\Gamma(\mathbf k,\mathbf p)
=\Gamma_{\rm dd}(\mathbf k-\mathbf p)
{\rm Sym}_{\mathbf k,\mathbf p}
\left\{
\varepsilon_{\mathbf k}^{(0)}
\left[1-2l_v(\mathbf k)\right]
\right\}
\end{equation}
accounts for the nonorthogonality of the correlated basis states,
and ${\rm Sym}_{\mathbf k,\mathbf p}\{F(\mathbf k,\mathbf p)\}
=F(\mathbf k,\mathbf p)+F(\mathbf p,\mathbf k)$ denotes symmetrization under
the exchange of $\mathbf k$ and $\mathbf p$. In the weak-pairing limit,
$V_\Gamma(\mathbf k,\mathbf p)=\left(|\varepsilon_{\mathbf k}^{(0)}|+|\varepsilon_{\mathbf p}^{(0)}|\right)\Gamma_{\rm dd}(\mathbf k-\mathbf p)$, in agreement with correlated-basis-function theory~\cite{Fan2015,FanKrotscheck2019}. In the ground state, $l_{v}(\mathbf{p})=(1-%
\mathcal{E}_{\mathbf{p}}/E_{\mathbf{p}})/2$ and $l_{u}(\mathbf{p})=\Delta _{%
\mathbf{p}}/(2E_{\mathbf{p}})$ with $E_{\mathbf{p}}=\sqrt{\mathcal{E}_{%
\mathbf{p}}^{2}+\left\vert \Delta _{\mathbf{p}}\right\vert ^{2}}$.
Consequently, Eq.~\eqref{Dk} becomes a self-consistent gap equation.

To solve Eq.~\eqref{Dk} together with the density constraint
$\int d^d k\, l_v(\mathbf k)/(2\pi)^d=\rho/\nu$, we use the imaginary-time
flow~\cite{ShiDemlerCirac2018}
\begin{equation}
\partial_\tau \mathcal G_{\mathbf k}
=
\{\mathcal H_{\mathbf k},\mathcal G_{\mathbf k}\}
-2\mathcal G_{\mathbf k}\mathcal H_{\mathbf k}\mathcal G_{\mathbf k},
\label{flow}
\end{equation}
for the covariance matrix
\begin{equation}
\mathcal G_{\mathbf k}
=
\begin{pmatrix}
1-l_v(\mathbf k) & l_u(\mathbf k) \\
l_u^\ast(\mathbf k) & l_v(\mathbf k)
\end{pmatrix},
\end{equation}
where the corresponding Bogoliubov-de Gennes Hamiltonian is
$\mathcal H_{\mathbf k}
=
\begin{pmatrix}
\mathcal E_{\mathbf k} & \Delta_{\mathbf k} \\
\Delta_{\mathbf k}^\ast & -\mathcal E_{\mathbf k}
\end{pmatrix}$.
The initial seed is chosen such that $l_v(\mathbf k)$ satisfies the density
constraint. During the flow, the time-dependent chemical potential $\mu(\tau)$
is adjusted to impose
$\int d^d k\,\partial_\tau l_v(\mathbf k)/(2\pi)^d=0$, so that the density is
conserved throughout the evolution~\cite{Shi2020}. At convergence,
$\partial_\tau\mathcal G_{\mathbf k}=0$, and the gap Eq.~\eqref{Dk} and
the density constraint are satisfied self-consistently.

\section{Universal interaction and correlated phase diagram}
\label{sec:molecules}

We now separate the many-body problem from the microscopic details of a particular molecule. Tight confinement along the \(z\) direction reduces the microwave-engineered interaction to a universal quasi-2D form. We first establish this reduction and then use the correlated theory of Sec.~II to determine the stability boundary and the $p$-wave pairing scale of the universal model.

\subsection{Universal interaction in two dimensions}
\label{subsec:universal_interaction}

For the microwave-shielded states considered here, the three-dimensional (3D) interaction can be written in the effective form~\cite{Deng2023,Deng2025Dual,Karman2025Double}
\begin{equation}
V_{\mathrm{3D}}(\mathbf{r})
=
\frac{C_{3}}{r^{3}}
\left(3\cos ^{2}\theta-1\right)
+
\frac{1}{r^{6}}
\sum_{l=0,2,4}C_{6l}Y_{l0}(\hat{r}),
\label{eq:V3D}
\end{equation}
where $\theta$ is the polar angle of $\mathbf r$ and $Y_{l0}(\hat r)$ is a spherical harmonic. We focus on $C_3>0$, for which the $r^{-3}$ tail is attractive in the $xy$ plane while the $r^{-6}$ terms generate the shielding core. The competition between these contributions produces an attractive minimum at $r=r_{\min}$.

We confine the molecules harmonically along $z$ with frequency $\omega_z$ and oscillator length $\sigma_z=\sqrt{\hbar/(M\omega_z)}$. Averaging Eq.~\eqref{eq:V3D} over the ground-state density along the $z$-direction gives the quasi-2D interaction
\begin{equation}
V_{\mathrm{2D}}(\rho)
=
\int_{-\infty}^{+\infty}dz\,
\frac{e^{-z^{2}/(2\sigma_z^{2})}}
{\sqrt{2\pi}\sigma_z}
V_{\mathrm{3D}}(\mathbf{r}),
\label{V2D}
\end{equation}
where $\rho$ is the in-plane intermolecular separation. The closed analytical expression used numerically is given in Appendix~\ref{2Dpotential}.

For sufficiently tight confinement along the $z$ direction,
$\sigma_z/r_{\min}\ll1$, the resulting interaction is accurately described
by the universal form
\begin{equation}
V_{\mathrm{eff}}(\rho)
=
\frac{C_{0}}{\sigma^{2}}
\left[
-\left(\frac{\sigma}{\rho}\right)^{3}
+\frac{1}{2}
\left(\frac{\sigma}{\rho}\right)^{6}
\right],
\label{eq:Veff}
\end{equation}
where $\sigma$ is the position of the minimum and $-C_0/(2\sigma^2)$ is its depth. Rescaling distance by $\sigma$ leaves a fixed shape, so the many-body problem depends only on $k_F\sigma$ and the dimensionless coupling $\bar C_0\equiv2MC_0/\hbar^2$. This collapse is useful because $\bar C_0$ and $k_F\sigma$ can be optimized without committing to a molecular species or a particular dressing protocol. We therefore determine the correlated phase diagram of Eq.~\eqref{eq:Veff} first and only afterwards search for microscopic $V_{\mathrm{2D}}(\rho)$ that reproduce the desired $(\bar C_0,\sigma)$. Section~IV shows that the relevant microscopic interactions closely follow Eq.~\eqref{eq:Veff}, validating this two-parameter reduction in the superfluid window.

\subsection{Phase diagram of the universal model}
\label{subsec:phase_diagram}

We consider a single-component gas ($\nu=1$), for which the leading odd-parity pairing channel is $p$ wave. Resolving the stability boundary requires particular care because the long-wavelength density response becomes singular at a spinodal instability~\cite{Krotscheck2000,Fan2015,FanKrotscheck2019}, signaling the onset of droplet formation. We therefore use a real-space infrared cutoff $R/\sigma=10^3$. Equation~\eqref{eq:FHNC_EL_main} is evaluated with a discrete Hankel transform~\cite{Johnson1987,Lemoine1994} on grids $k_n=\alpha_n/R$ and $\rho_n=\alpha_n/K$, where $\alpha_n$ is the $n$th zero of $J_0$ and $K=k_N$. We use $N=3\times10^4$ points, which resolves distances down to $\rho_1/\sigma\simeq0.026$, and have checked convergence against larger infrared and ultraviolet cutoffs. The flow Eq.~\eqref{flow} is solved on a logarithmic momentum grid $k_n/k_F=1\pm D\Lambda^{n-N_c}$ concentrated around the Fermi surface, with $\Lambda=1.05$, $D\simeq49$, and $D\Lambda^{-N_c}\simeq10^{-3}$. These choices give converged gaps throughout the stable region.

\begin{figure}[t]
    \centering
    \includegraphics[width=\columnwidth]{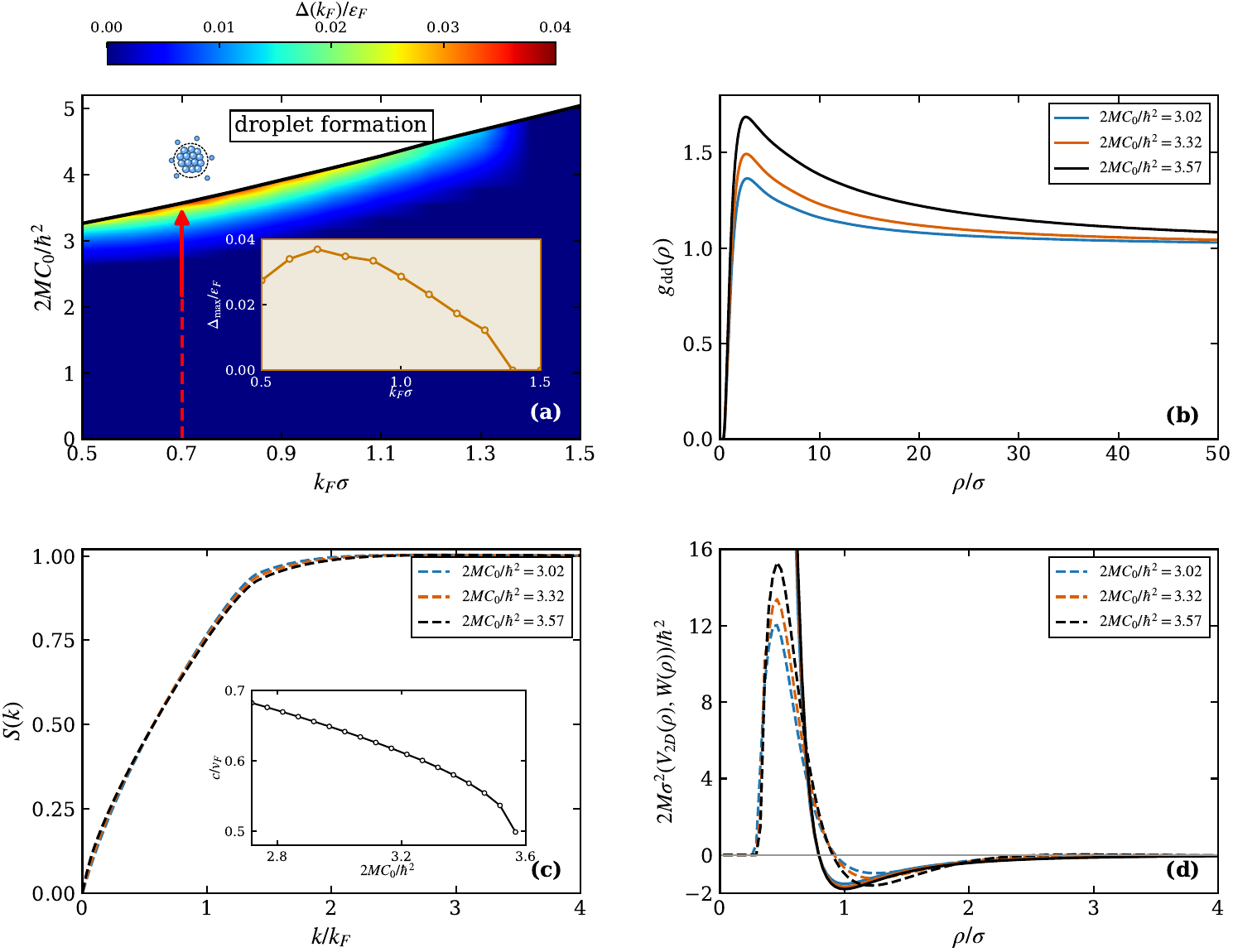}
    \caption{
    Stability, correlation functions, structure factors, many-body effective potentials, and $p$-wave superfluid gaps in a
    single-component ($\nu=1$) 2D molecular gas with $V_{\mathrm{eff}}(\rho)$.
    (a) Phase diagram in the
    $k_F\sigma$--$\bar{C}_0$ plane.
    The solid black curve marks the long-wavelength spinodal boundary; above it, the homogeneous fluid is unstable toward density collapse associated with droplet formation.
    The color map shows the superfluid gap
    $\Delta(k_F)/\varepsilon_F$ in the stable region.
    The inset shows the maximum gap
    $\Delta_{\max}(k_F)/\varepsilon_F$ at each density.
    The red arrow indicates the cut $k_F\sigma=0.7$ used in
    panels (b)--(d).
    (b) Dynamical correlation function $g_{\mathrm{dd}}(\rho)$ at
    $k_F\sigma=0.7$ for
    $\bar{C}_0=3.02$, $3.32$, and $3.57$
    (blue, red, and black, respectively).
    (c) Corresponding static structure factor $S(k)$.
    The inset shows the sound velocity $c/v_F$ as a function of
    $\bar{C}_0$ along the same $k_F\sigma=0.7$ cut.
    (d) Bare two-body interaction $V_{\mathrm{eff}}(\rho)$
    (solid curves) and the correlation-renormalized interaction
    $W(\rho)$ screened by particle-hole excitations
    (dashed curves), for the same parameters and color convention
    as in panels (b) and (c).
    }
    \label{phaseC3C6}
\end{figure}

Figure~\ref{phaseC3C6}(a) shows the resulting phase diagram in the $k_F\sigma$--$\bar C_0$ plane. The black curve is the long-wavelength spinodal boundary of the homogeneous fluid. Although our calculation does not assume the structure of the inhomogeneous state beyond this boundary, the instability is naturally associated with density collapse and droplet formation. The cut at $k_F\sigma=0.7$ illustrates how the instability develops. As $\bar C_0$ increases, $g_{\mathrm{dd}}(\rho)$ acquires a pronounced peak
just outside the shielding core [Fig.~\ref{phaseC3C6}(b)], reflecting enhanced
correlations at separations near the attractive minimum. Simultaneously, the small-$k$ slope of $S(k)$ increases [Fig.~\ref{phaseC3C6}(c)], so the sound velocity decreases and vanishes at the spinodal boundary. The instability is therefore visible directly in the optimized density correlations rather than imposed as an external criterion.

A notable feature is that increasing density can stabilize the 2D fluid, opposite to the dilute 3D trend. This behavior follows from the long-wavelength stability condition $D(\mathbf 0)>0$ in Eq.~\eqref{eq:FHNC_EL_main}, where
\begin{equation}
D(\mathbf{0})
=
1+
2c_F^{2}
\frac{\rho}{\varepsilon_F}
\int d^{d}r\,\bar{V}(\mathbf r).
\end{equation}
For a single-component gas in 3D, $\rho/\varepsilon_F=Mk_F/(3\pi^2)$ grows with density, thus, an attractive contribution to $\bar V$ is therefore amplified algebraically and drives $D(0)$ toward zero. In 2D, by contrast, $\rho/\varepsilon_F=M/(2\pi)$ is density independent. The residual density dependence is then carried by the correlation-induced interaction $\bar{V}(\rho)$. Increasing $k_F$ enhances the Fermi pressure and suppresses the accumulation
of molecules near the attractive minimum, thereby reducing the peak in
$g_{\mathrm{dd}}(\rho)$. This, in turn, weakens the attractive contribution
$V(\rho)g_{\mathrm{dd}}(\rho)$ entering Eq.~\eqref{Vph}, making the
correlation-renormalized interaction $\bar V(\rho)$ less attractive and stabilizing
the homogeneous 2D fluid.

The same correlations strongly renormalize the interaction relevant for pairing. Figure~\ref{phaseC3C6}(d) compares the bare $V_{\mathrm{eff}}(\rho)$ with $W(\rho)$, which includes the correlated medium response. The divergent repulsive core of the bare potential is replaced by a finite short-range maximum because the correlation function $g_\mathrm{dd}$ is strongly suppressed for $\rho<\sigma$. The attractive minimum is also shifted upward and toward larger $\rho$.
As $\bar{C}_0$ increases, both the magnitude of the long-range attraction and the height of the short-range maximum in $W(\rho)$ increase. Thus, when the mean interparticle distance becomes comparable to the size of
the shielding core, neither the short-range repulsion nor the attractive
interaction relevant for pairing is faithfully represented by the bare
potential. The attractive
part of $W(\rho)$ provides the dominant interaction in the $p$-wave pairing channel.

For the single-component case, $\nu=1$, we focus on the chiral $p$-wave
solution
$\Delta_{\mathbf{k}}=\Delta(k)e^{\pm i\varphi_{\mathbf{k}}}$.
The color scale in Fig.~\ref{phaseC3C6}(a) shows the superfluid gap
$\Delta(k_F)/\varepsilon_F$ throughout the stable homogeneous region.
At fixed density, the gap increases with interaction strength and reaches
its maximum at the spinodal boundary approached from the stable side.
We denote this maximal gap at a given density by $\Delta_{\max}$. As shown in the inset of Fig.~\ref{phaseC3C6}(a),
$\Delta_{\max}/\varepsilon_F$ depends nonmonotonically on density.
This behavior originates from the competition between the long-range
attraction and the short-range shielding core. At low density, the mean
interparticle spacing is much larger than the shielding-core size, so that
the molecules predominantly sample the attractive $-1/\rho^3$ tail.
Increasing $k_F$ therefore strengthens the effective attraction and enhances
the pairing gap. At higher density, however, the mean interparticle spacing
becomes comparable to the shielding-core radius, and the molecules
increasingly probe the short-range repulsive part of the interaction.
The resulting suppression of the effective attraction reduces the gap,
which eventually vanishes near $k_F\sigma=1.4$. The global maximum is
$\Delta_{\max}/\varepsilon_F=3.7\times10^{-2}$,
obtained at $k_F\sigma=0.7$ and $\bar C_0=3.567$.
This optimum motivates the interaction window considered below when mapping
the universal phase diagram onto microscopic microwave-shielding schemes.

\section{$p$-wave superfluidity in MSPM{\lowercase{s}}}
\label{subsec:shielding_realization}

We now translate the universal phase diagram into microscopic molecular
parameters. The favorable pairing window lies at moderate coupling,
$\bar C_0\simeq3$--$5$, rather than at the strongest attraction available
from microwave dressing. For a microscopic quasi-2D interaction
$V_{\mathrm{2D}}(\rho)$, let $V_{\min}<0$ denote its minimum and $\sigma$
the position of that minimum. In the tight-confinement limit,
$V_{\mathrm{2D}}$ reduces to the universal interaction
$V_{\mathrm{eff}}$, and matching to Eq.~\eqref{eq:Veff} gives
$C_0=2|V_{\min}|\sigma^2$, or equivalently
$\bar C_0=4M|V_{\min}|\sigma^2/\hbar^2$.
This mapping provides a direct criterion for selecting the microwave and
confinement parameters. We first use it to assess single-microwave shielding
and then construct dual-microwave working regimes that balance pairing,
many-body stability, collisional loss, confinement, and robustness against
microwave-amplitude fluctuations.

\begin{figure}[t]
\centering
\includegraphics[width=\columnwidth]{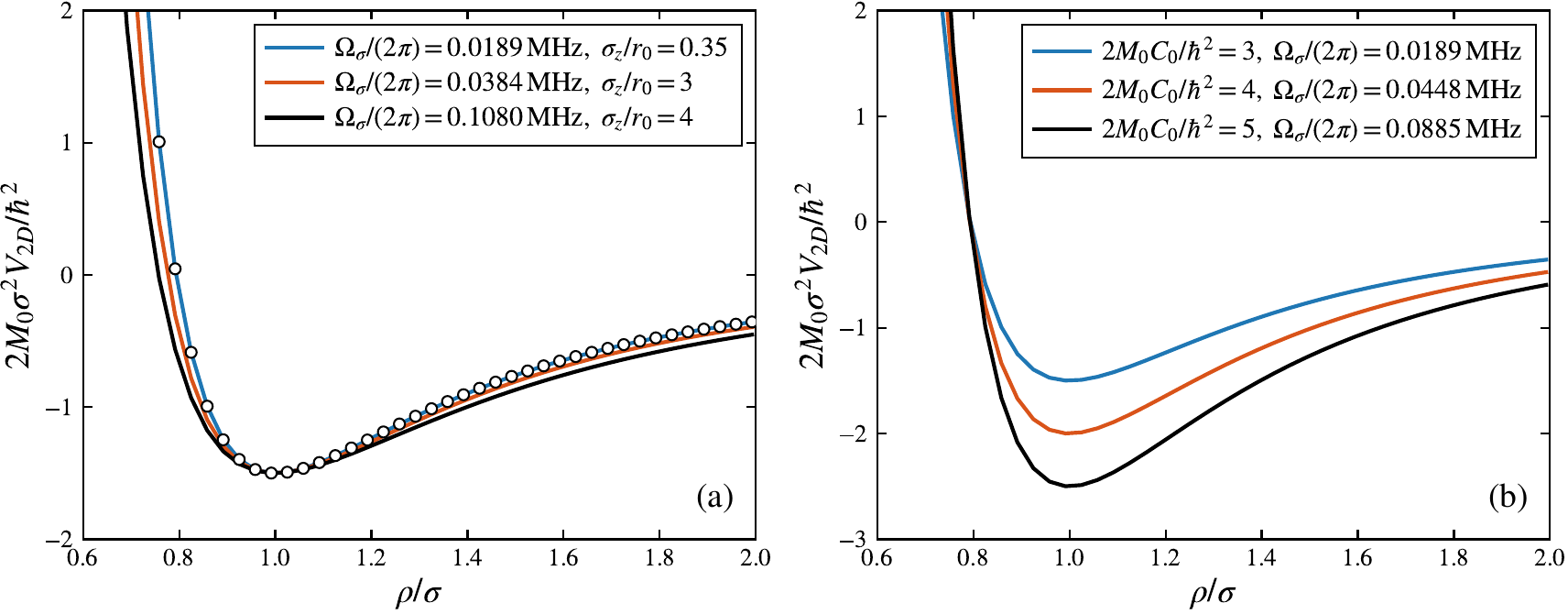}
\caption{
Effective quasi-2D interactions for single-microwave shielding.
The solid curves show the quasi-2D interaction $V_{\mathrm{2D}}(\rho)$ obtained from Eq.~\eqref{V2D}. Here, $\sigma$ denotes the position of the minimum of $V_{\mathrm{2D}}(\rho)$, and $r_0=10^3a_0$.
(a) Realization of $V_{\mathrm{2D}}(\rho)$ with $\bar{C}_0=3$ for LiRb molecules by varying the microwave Rabi frequency and the confinement length, while fixing the relative detuning at $\delta_\sigma=\Delta_\sigma/\Omega_\sigma=1$. The open circles show the corresponding effective potential $V_{\mathrm{eff}}(\rho)$ with $\bar{C}_0=3$.
(b) Tuning $V_{\mathrm{2D}}(\rho)$ from $\bar{C}_0=3$ to $\bar{C}_0=5$ for LiRb molecules by varying the microwave Rabi frequency at fixed $\sigma_z/r_0=0.35$ and $\delta_\sigma=1$.
}
\label{figpotential}
\end{figure}

\begin{table}[t]
\caption{Microwave and confinement parameters for realizing $\bar C_0=3$, $4$, and $5$ with single-microwave shielding. The relative detuning is fixed at $\delta_\sigma=\Delta_\sigma/\Omega_\sigma=1$. The quantities $r_{\min}$ and $\sigma$ are the positions of the minima of the 3D and quasi-2D interactions, respectively.}
\label{tab:single_mw_parameters}
\centering
\small
\setlength{\tabcolsep}{4.5pt}
\renewcommand{\arraystretch}{1.08}
\begin{tabular}{ccccc}
\toprule
Quantity & $\bar C_0$ & NaK & LiRb & KAg \\
\midrule
$\Omega_\sigma/(2\pi)$ (MHz) & 3 & $0.355$ & $0.0189$ & $2.8\times10^{-4}$ \\
& 4 & $0.864$ & $0.0448$ & $6.7\times10^{-4}$ \\
& 5 & $1.735$ & $0.0885$ & $1.3\times10^{-3}$ \\
\addlinespace
$r_{\min}/r_0$ & 3 & $3.55$ & $12.6$ & $82.13$ \\
& 4 & $2.64$ & $9.45$ & $61.36$ \\
& 5 & $2.09$ & $7.53$ & $49.11$ \\
\addlinespace
$\sigma_z/r_0$ & $3,4,5$ & $0.20$ & $0.35$ & $1.00$ \\
\addlinespace
$\sigma/r_0$ & 3 & $3.61$ & $12.8$ & $83.48$ \\
& 4 & $2.69$ & $9.61$ & $62.37$ \\
& 5 & $2.14$ & $7.66$ & $49.9$ \\
\bottomrule
\end{tabular}
\end{table}

\subsection{Single-microwave shielding}
\label{subsec:SMS}

We first consider the single-microwave shielding scheme, in which the
molecules are dressed by a single circularly ($\sigma$-) polarized microwave
field. Figure~\ref{figpotential}(a) uses LiRb as a representative example to
examine when the microscopic quasi-2D interaction is faithfully described by
the universal form. The blue, red, and black curves correspond to confinement
lengths $\sigma_z/r_0=0.35$, $3$, and $4$, respectively, with the microwave
parameters adjusted to realize the same $\bar C_0=3$. At fixed relative
detuning $\delta_\sigma=\Delta_\sigma/\Omega_\sigma=1$, the corresponding Rabi
frequencies are $\Omega_\sigma/(2\pi)=1.89\times10^{-2}$,
$3.84\times10^{-2}$, and $1.08\times10^{-1}\,\mathrm{MHz}$.
The positions of the minima of the 3D and quasi-2D
potentials are $(r_{\min},\sigma)/r_0=(12.6,12.8)$, $(9.95,10.75)$, and
$(7.05,8.28)$, respectively. Here, $r_0=10^3a_0$, with $a_0$ the Bohr radius.

Weakening the confinement along the $z$ direction has two linked consequences.
Averaging over finite $z$, where the long-range attraction is reduced, weakens
the effective in-plane attraction, so a larger $\Omega_\sigma$ is required to
retain the same $\bar C_0$. The larger Rabi frequency in turn decreases
$r_{\min}$, making the confinement less tight relative to the shielding core.
The resulting finite-$z$ averaging also softens the short-range repulsion. Consequently,
the microscopic system can become spinodally unstable at a smaller $\bar C_0$
than predicted by the universal model, thereby reducing the maximum accessible
superfluid gap. Sufficiently tight confinement avoids this problem. For
$\sigma_z/r_0=0.35$, the microscopic $V_{\mathrm{2D}}$ and the universal
$V_{\mathrm{eff}}$ in Fig.~\ref{figpotential}(a) are nearly indistinguishable
over the region relevant to the many-body physics.

Figure~\ref{figpotential}(b) and Table~\ref{tab:single_mw_parameters}
show how $\bar C_0=3$, $4$, and $5$ can be realized for NaK, LiRb,
and KAg. These working points expose the central limitation of the
single-microwave scheme: obtaining only moderate attraction requires a
small $\Omega_\sigma$, but reducing the Rabi frequency also decreases
the separation between dressed collision channels and weakens collisional
shielding~\cite{KarmanHutson2018,KarmanHutson2019,Schindewolf2022,
Deng2023,Stevenson2024}. The attractive tail and the repulsive shielding
core originate from the same dressing field and therefore cannot be tuned
independently. Reducing the attraction thus inevitably compromises
collisional protection, motivating the dual-microwave scheme considered
next.

\subsection{Dual-microwave shielding}
\label{subsec:DMS}
Dual-microwave shielding supplies the missing control knob. A circularly
($\sigma$-) polarized field provides the main collisional shielding, while an
additional linearly ($\pi$-) polarized field compensates the long-range
attraction while retaining the intermediate-range shielding
barrier~\cite{Deng2025Dual,Karman2025Double,Bigagli2024,Yuan2025Loss,
Schindewolf2026RMP}. The two Rabi frequencies and two detunings define a
four-dimensional control space, but the correlated phase diagram sharply
restricts the useful part of this space. Efficient shielding favors a large
$\Omega_\sigma$, whereas stability against droplet formation, feasible
confinement, and robustness against microwave fluctuations limit how large
$\Omega_\sigma$ can be.

\begin{figure}[t]
    \centering
    \includegraphics[width=\columnwidth]{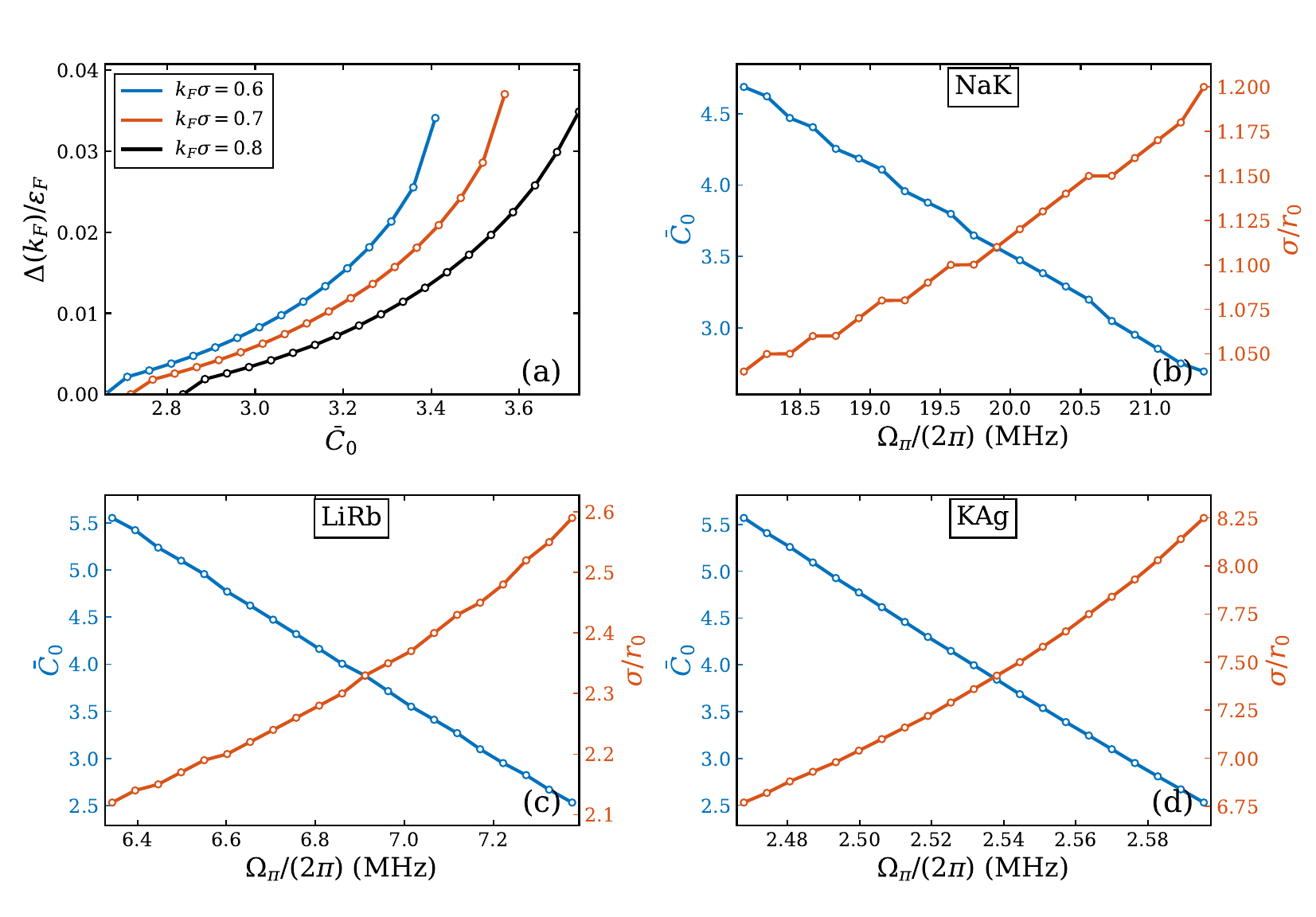}
    \caption{$p$-wave superfluid gap and microwave-controlled interaction parameters.
(a) $p$-wave superfluid gap $\Delta(k_F)/\varepsilon_F$ as a function of the dimensionless interaction strength $\bar{C}_0$ for the effective potential $V_{\mathrm{eff}}$, shown for $k_F\sigma=0.6$, $0.7$, and $0.8$. (b)--(d) Dimensionless interaction strength $\bar{C}_0$ and interaction length scale $\sigma/r_0$ as functions of the $\pi$-field Rabi frequency $\Omega_\pi/(2\pi)$ for NaK, LiRb, and KAg, respectively.}
    \label{gapC0sigma}
\end{figure}

First, increasing $\Omega_\sigma$ reduces the size of the 3D
shielding core $r_{\min}$. Avoiding the spinodal instability then requires
correspondingly tighter confinement along the $z$ direction, typically
$\sigma_z\sim0.2\,r_{\min}$. Such a small $\sigma_z$ translates into a large
trapping frequency, which may become experimentally
demanding for relatively light molecules such as NaK, even when using a single
minimum of a deep optical lattice. Second, a large $\Omega_\sigma$ also
produces a stronger attractive background interaction. To recover the
moderate effective attraction, $\bar C_0\sim3$, required for an appreciable
superfluid gap, this attraction must be compensated by a $\pi$-polarized
microwave with a correspondingly large Rabi frequency $\Omega_\pi$. This
compensation must be sufficiently precise because the pairing gap is highly
sensitive to $\bar C_0$ in this regime. As shown in
Fig.~\ref{gapC0sigma}(a), decreasing $\bar C_0$ by only $0.1$ from its
boundary value $\bar C_0=3.57$ reduces the maximal gap from
$\Delta_{\max}/\varepsilon_F\simeq3.7\times10^{-2}$ to
$2.4\times10^{-2}$, while the gap eventually vanishes around
$\bar C_0=2.7$. The superfluid state is therefore particularly sensitive to
fluctuations of the $\pi$-microwave Rabi frequency when the system operates
near $\bar C_0\sim3$.

A larger attractive background generated by the $\sigma$ field also requires a stronger $\pi$ field for compensation. The useful interval in $\bar C_0$ is narrow, so amplitude fluctuations of the compensating field are converted directly into fluctuations of the pairing interaction. Rather than assign a universal experimental noise level, we quantify this conversion through a local sensitivity and use it as an explicit optimization constraint. The preferred $\Omega_\sigma$ is therefore intermediate: large enough for robust collisional shielding, but not so large that the required confinement or $\pi$-field compensation becomes impractically stringent. The relevance of microwave polarization and dressed-state control to shielding has been characterized experimentally and theoretically in Refs.~\cite{Deng2025Dual,Zhang2024Dressed,Karman2025Double,Shi2026NaRb}.

To quantify the sensitivity of the effective interaction to fluctuations of the
microwave amplitude, we introduce the relative detuning
$\delta_\pi=\Delta_\pi/\Omega_\sigma$ and the relative Rabi frequency
$\bar{\Omega}_\pi=\Omega_\pi/\Omega_\sigma$ for the $\pi$-polarized field.
We focus on the regime $\delta_\sigma=\Delta_\sigma/\Omega_\sigma\sim1$.
A substantially larger $\delta_\sigma$ weakens the microwave dressing and hence
the shielding, whereas a smaller $\delta_\sigma$ produces a deeper interaction
potential and increases its sensitivity to fluctuations of the microwave
parameters. This consideration substantially restricts the parameter regime
relevant for optimizing the superfluid gap.

The dimensionless interaction strength can be expressed in terms of the
long-range coefficient $C_3$ as
\begin{equation}
\bar{C}_0
=
\frac{2MC_3}{\hbar^2\sigma}
=
\frac{\gamma_0}{3\bar{\sigma}}
\left(
\left|\langle +|I_\sigma|+\rangle\right|^2
-
2\left|\langle +|I_\pi|+\rangle\right|^2
\right),
\label{eq:Cbar_dressed}
\end{equation}
where $\bar{\sigma}=\sigma/r_0$ and
$\gamma_0=Md^2/(2\pi\varepsilon_0\hbar^2r_0)$
is a dimensionless parameter determined by the molecular species.
The highest dressed state $|+\rangle$ is an eigenstate of the
single-molecule Hamiltonian
$H_{\mathrm{s}}=\Omega_\sigma h_{\mathrm{s}}$, with
\begin{equation}
h_{\mathrm{s}}
=
\begin{pmatrix}
0 & \frac{1}{2} & \frac{1}{2}\bar{\Omega}_\pi \\
\frac{1}{2} & -\delta_\sigma & 0 \\
\frac{1}{2}\bar{\Omega}_\pi & 0 & -\delta_\pi
\end{pmatrix},
\label{eq:single_molecule_h}
\end{equation}
written in the basis
$\{J,M\}=\{|0,0\rangle,|1,1\rangle,|1,0\rangle\}$ of rotational states.
Here, $I_\sigma=|1,1\rangle\langle0,0|$ and
$I_\pi=|1,0\rangle\langle0,0|$ describe the transitions induced by $\sigma$- and $\pi$-polarized fields,
respectively. In terms of the Euler angles $\alpha$ and $\beta$,
$|+\rangle
=
(\cos\alpha,\sin\alpha\cos\beta,\sin\alpha\sin\beta)^{\mathrm{T}}$, and Eq.~\eqref{eq:Cbar_dressed} becomes
\begin{equation}
\bar{C}_0
=
\frac{\gamma_0}{6\bar{\sigma}}
\sin^2\alpha\cos^2\alpha
\left(3\cos2\beta-1\right).
\label{Cb0}
\end{equation}
This expression separates the molecular length scale, through
$\gamma_0/\bar{\sigma}$, from the microwave dressing encoded in the
angles $\alpha$ and $\beta$.

We next quantify the sensitivity of $\bar{C}_0$ to fluctuations of the
$\pi$-field amplitude. Numerically, we find that, for small relative Rabi-frequency fluctuations
$\delta\bar{\Omega}_\pi/\bar{\Omega}_\pi$, the induced variation of
$\bar{\sigma}$ is negligible compared with that of $C_3$. We therefore define the sensitivity
\begin{equation}
s_0
\equiv
\frac{\partial\bar{C}_0}{\partial l_\pi}\sim\frac{\gamma_0}{3\bar{\sigma}}
\frac{\partial}{\partial l_\pi}
\left[
\left|\langle +|I_\sigma|+\rangle\right|^2
-
2\left|\langle +|I_\pi|+\rangle\right|^2
\right],
\label{dCl}
\end{equation}
where $l_\pi=\ln\bar{\Omega}_\pi$. For the parameter regime considered here, $s_0<0$, because increasing
$\bar{\Omega}_\pi$ strengthens the compensating contribution of the
$\pi$ field and therefore reduces the net attractive interaction. Equation~\eqref{dCl} can be evaluated analytically. Choosing real $|+\rangle$ such
that $\langle+|\partial_{l_\pi}|+\rangle=0$,
one obtains, for $|n\rangle\neq|+\rangle$,
\begin{equation}
\left\langle n
\middle|
\frac{\partial}{\partial l_\pi}
\middle|
+\right\rangle
=
\frac{
\bar{\Omega}_\pi
\langle n|
(I_\pi+I_\pi^\dagger)
|+\rangle
}{
2(e_+-e_n)
},
\label{eq:eigenstate_derivative}
\end{equation}
where $e_+$ and $e_n$ are the corresponding eigenvalues of
$h_{\mathrm{s}}$.

At a working point with a prescribed value of $\bar{C}_0$,
Eq.~\eqref{dCl} can equivalently be written as
\begin{equation}
s_0
=
\bar{C}_0
\frac{\partial}{\partial l_\pi}
\ln
\left[
\left|\langle +|I_\sigma|+\rangle\right|^2
-
2\left|\langle +|I_\pi|+\rangle\right|^2
\right].
\label{s0}
\end{equation}
For a small fluctuation of the $\pi$-field Rabi frequency,
\begin{equation}
\delta\bar{C}_0
\simeq
s_0\,
\frac{\delta\bar{\Omega}_\pi}{\bar{\Omega}_\pi}.
\label{eq:C_fluctuation}
\end{equation}
Consequently, requiring
$|\delta\bar{C}_0|<\delta_C$ imposes the bound
\begin{equation}
|s_0|
<
\frac{\delta_C}
{\delta\bar{\Omega}_\pi/\bar{\Omega}_\pi}.
\label{eq:sensitivity_bound}
\end{equation}
For $\delta_C\simeq0.1$, a relative Rabi-frequency fluctuation of $1\%$
requires $|s_0|\lesssim10$, whereas improving the fluctuation to $0.1\%$
relaxes this bound to $|s_0|\lesssim100$. These estimates provide a direct
connection between experimentally achievable microwave stability and the
range of interaction parameters that can be used for robust superfluidity.

We can therefore determine the microwave parameters without performing a
brute-force search over the full four-dimensional parameter space. The
procedure is as follows. First, we choose $\delta_\sigma\sim1$ and consider
$\delta_\pi>\delta_\sigma$. For each chosen $\delta_\pi$, we specify an
experimentally acceptable sensitivity $|s_0|$ and solve Eq.~\eqref{s0} for
$\bar{\Omega}_\pi$ at the target interaction strength, e.g.,
$\bar C_0=3$. Second, for a trial absolute Rabi frequency $\Omega_\sigma$, we determine the
3D shielding-core size $r_{\min}$ and choose the confinement
length $\sigma_z\simeq0.2\,r_{\min}$. We then calculate the corresponding
quasi-2D interaction $V_{2D}(\rho)$ and extract its interaction strength
$\bar C_0$ and characteristic length $\sigma$. By varying $\Omega_\sigma$, we
identify the working point at which the microscopic interaction realizes the
target $\bar C_0$. The complete set of microwave parameters is then
$(\Omega_\sigma,
\Delta_\sigma=\Omega_\sigma\delta_\sigma,\Omega_\pi=\Omega_\sigma\bar{\Omega}_\pi,
\Delta_\pi=\Omega_\sigma\delta_\pi)$.

Finally, we evaluate the elastic and inelastic scattering rates using a
coupled-channel calculation at a collision energy near the Fermi energy
$\varepsilon_F=\hbar^2k_F^2/(2M)$, taking $k_F\sigma\simeq0.7$.
As practical design targets, we seek an inelastic rate coefficient
$\beta_{\mathrm{inel}}\lesssim10^{-13}\,\mathrm{cm^3\,s^{-1}}$ and an
elastic-to-inelastic (``good-to-bad'') ratio
$\gamma_{\mathrm{gb}}\gtrsim5\times10^3$. If these targets are not met,
we allow a larger microwave sensitivity $|s_0|$, within the experimental
tolerance set by Eq.~\eqref{eq:sensitivity_bound}. This permits a larger
$\Omega_\sigma$ and hence stronger collisional shielding. The optimization
therefore balances loss suppression, microwave robustness, and sufficient
attraction for pairing without crossing the spinodal boundary. We additionally
require the confinement along the $z$ direction to remain experimentally
plausible. As a practical criterion, we require
$\omega_z/(2\pi)\lesssim320~\mathrm{kHz}$, which represents an experimentally
challenging but state-of-the-art scale for tight optical-lattice confinement.

\begin{table}[t]
\caption{Representative dual-microwave working points for NaK, LiRb, and KAg. Here $r_{\min}$ and $\sigma$ denote the minima of the 3D and quasi-2D interactions, $\omega_z$ is the confinement frequency, and $T_F$ is evaluated at $k_F\sigma=0.7$. The sensitivity $s_0$ is defined in Eq.~\eqref{dCl}; $\beta_{\mathrm{inel}}$ and $\gamma_{\mathrm{gb}}$ are the inelastic two-body rate coefficient and elastic-to-inelastic scattering ratio, respectively.}
\label{tab:dual_mw_parameters}
\centering
\small
\setlength{\tabcolsep}{5.5pt}
\renewcommand{\arraystretch}{1.12}
\begin{tabular}{cccc}
\toprule
Quantity & NaK & LiRb & KAg \\
\midrule
$\Omega_\sigma/(2\pi)$ (MHz) & $32.8$ & $10$ & $3.36$ \\
$\Delta_\sigma/(2\pi)$ (MHz) & $32.8$ & $12$ & $3.36$ \\
$\Omega_\pi/(2\pi)$ (MHz) & $20.8$ & $7.2$ & $2.57$ \\
$\Delta_\pi/(2\pi)$ (MHz) & $39.4$ & $15$ & $4.04$ \\
$r_{\min}/r_0$ & $1.12$ & $2.42$ & $7.77$ \\
$\sigma_z/r_0$ & $0.20$ & $0.35$ & $1.00$ \\
$\sigma/r_0$ & $1.16$ & $2.47$ & $7.9$ \\
$\omega_z/(2\pi)$ (kHz) & $1428$ & $316$ & $24$ \\
$T_F$ (nK) & $500$ & $75$ & $4.6$ \\
$|s_0|$ & $19$ & $20$ & $60$ \\
$\beta_{\mathrm{inel}}$ ($\mathrm{cm^3\,s^{-1}}$) & $8.5\times10^{-13}$ & $4.8\times10^{-14}$ & $5.3\times10^{-16}$ \\
$\gamma_{\mathrm{gb}}$ & $238$ & $5147$ & $1.05\times10^6$ \\
\bottomrule
\end{tabular}
\end{table}

Using the procedure described above, we identify representative microwave and confinement parameters for different molecular species, as summarized in Table~\ref{tab:dual_mw_parameters}. For each molecule, we first choose $(\Omega_\sigma,\Delta_\sigma,\Delta_\pi)$ and the confinement length $\sigma_z$, and then determine the $\pi$-field Rabi frequency $\Omega_\pi$ required to realize the target interaction strength $\bar{C}_0=3$. The corresponding three- and two-dimensional interaction minima, $r_{\min}$ and $\sigma$, the confinement frequency $\omega_z$, the Fermi temperature $T_F=\varepsilon_F/k_B$ at $k_F\sigma=0.7$, the sensitivity $|s_0|$, the inelastic scattering rate coefficient $\beta_{\mathrm{inel}}$, and the good-to-bad scattering ratio $\gamma_{\mathrm{gb}}$ are also listed. To illustrate both the tunability of the interaction and its sensitivity to the $\pi$-field amplitude, we plot $\bar{C}_0$ and $\sigma/r_0$ as functions of $\Omega_\pi$ for NaK, LiRb, and KAg in Figs.~\ref{gapC0sigma}(b)--\ref{gapC0sigma}(d), respectively.

For NaK, the relatively small molecular mass and dipole moment force the
system into a demanding parameter regime. Achieving sufficiently strong
shielding requires a comparatively large Rabi frequency,
$\Omega_\sigma/(2\pi)=32.8\,\mathrm{MHz}$, which yields a small shielding
core, $r_{\min}/r_0=1.12$. Stabilizing the homogeneous fluid against the
spinodal instability then requires $\sigma_z/r_0\simeq0.20$, corresponding to
$\omega_z/(2\pi)\simeq1.43\,\mathrm{MHz}$, well above the confinement scale
adopted in our practical criterion. At $k_F\sigma=0.7$, where
$T_F\simeq500\,\mathrm{nK}$, the same working point gives
$\beta_{\mathrm{inel}}=8.5\times10^{-13}\,\mathrm{cm^3\,s^{-1}}$ and
$\gamma_{\mathrm{gb}}=238$. Increasing $\Omega_\sigma$ would improve
collisional shielding, but would further shrink the shielding core and demand
still tighter confinement. Thus, within the constraints considered here,
simultaneously achieving low loss, stability against droplet formation, and
experimentally feasible confinement is particularly challenging for NaK.

LiRb shifts the balance into a substantially more favorable regime. With
$M=93\,\mathrm{u}$ and $d=4.2\,\mathrm{D}$, a smaller Rabi frequency,
$\Omega_\sigma/(2\pi)=10\,\mathrm{MHz}$, is sufficient to provide strong
shielding, yielding $r_{\min}/r_0=2.42$. The choice
$\sigma_z/r_0=0.35$ corresponds to
$\omega_z/(2\pi)=316\,\mathrm{kHz}$, which lies within the practical
confinement range adopted here and remains far below the requirement for NaK.
At $k_F\sigma=0.7$, the Fermi temperature is $T_F=75\,\mathrm{nK}$, while
the coupled-channel calculation gives
$\beta_{\mathrm{inel}}=4.8\times10^{-14}\,\mathrm{cm^3\,s^{-1}}$ and
$\gamma_{\mathrm{gb}}=5147$. The sensitivity $s_0=-20$ implies that a
$0.5\%$ relative fluctuation $\delta\Omega_\pi/\Omega_\pi$ changes
$\bar C_0$ by approximately $0.1$. LiRb therefore provides a favorable
compromise among collisional shielding, experimentally accessible confinement,
and robustness against microwave-amplitude fluctuations.

KAg illustrates the opposite limit. Its larger mass, $M=147\,\mathrm{u}$, and dipole moment, $d=8.5\,\mathrm{D}$, allow strong shielding already at $\Omega_\sigma/(2\pi)=3.36\,\mathrm{MHz}$. The resulting core is large, with $r_{\min}/r_0=7.77$ and $\sigma/r_0=7.9$, so that $\sigma_z/r_0=1$ requires only $\omega_z/(2\pi)=24\,\mathrm{kHz}$. The calculated loss figures are correspondingly favorable: at $k_F\sigma=0.7$, $\beta_{\mathrm{inel}}=5.3\times10^{-16}\,\mathrm{cm^3\,s^{-1}}$ and $\gamma_{\mathrm{gb}}=1.05\times10^6$. The price is twofold. The large length scale lowers the Fermi temperature to $T_F=4.6\,\mathrm{nK}$, imposing more stringent cooling requirements, while the effective attraction becomes more sensitive to the compensating field. With $s_0=-60$, a relative fluctuation of $0.17\%$ in $\Omega_\pi$ changes $\bar C_0$ by approximately $0.1$. KAg therefore offers exceptional collisional protection at the cost of a lower operating temperature and stricter microwave-amplitude stability.

The comparison in Table~\ref{tab:dual_mw_parameters} therefore exposes a systematic species trade-off. Larger mass and dipole moment enlarge the shielding core and suppress inelastic scattering, substantially relaxing the required confinement along the $z$ direction. At the same time, the target density corresponds to a lower $T_F$, while the compensated interaction becomes more sensitive to the microwave parameters. The most favorable species is therefore not simply the most strongly dipolar one, but rather the one that best balances collisional protection, attainable temperature, confinement, and microwave-control stability.

\subsection{Microscopic phase diagrams for LiRb and KAg}
\label{subsec:phaseMSPMs}

For NaK, simultaneously achieving sufficiently low collisional loss and experimentally feasible confinement along the $z$ direction is particularly challenging. We therefore compute the full microscopic phase diagrams for the two more favorable examples, LiRb and KAg.

\begin{figure}[t]
    \centering
    \includegraphics[width=\columnwidth]{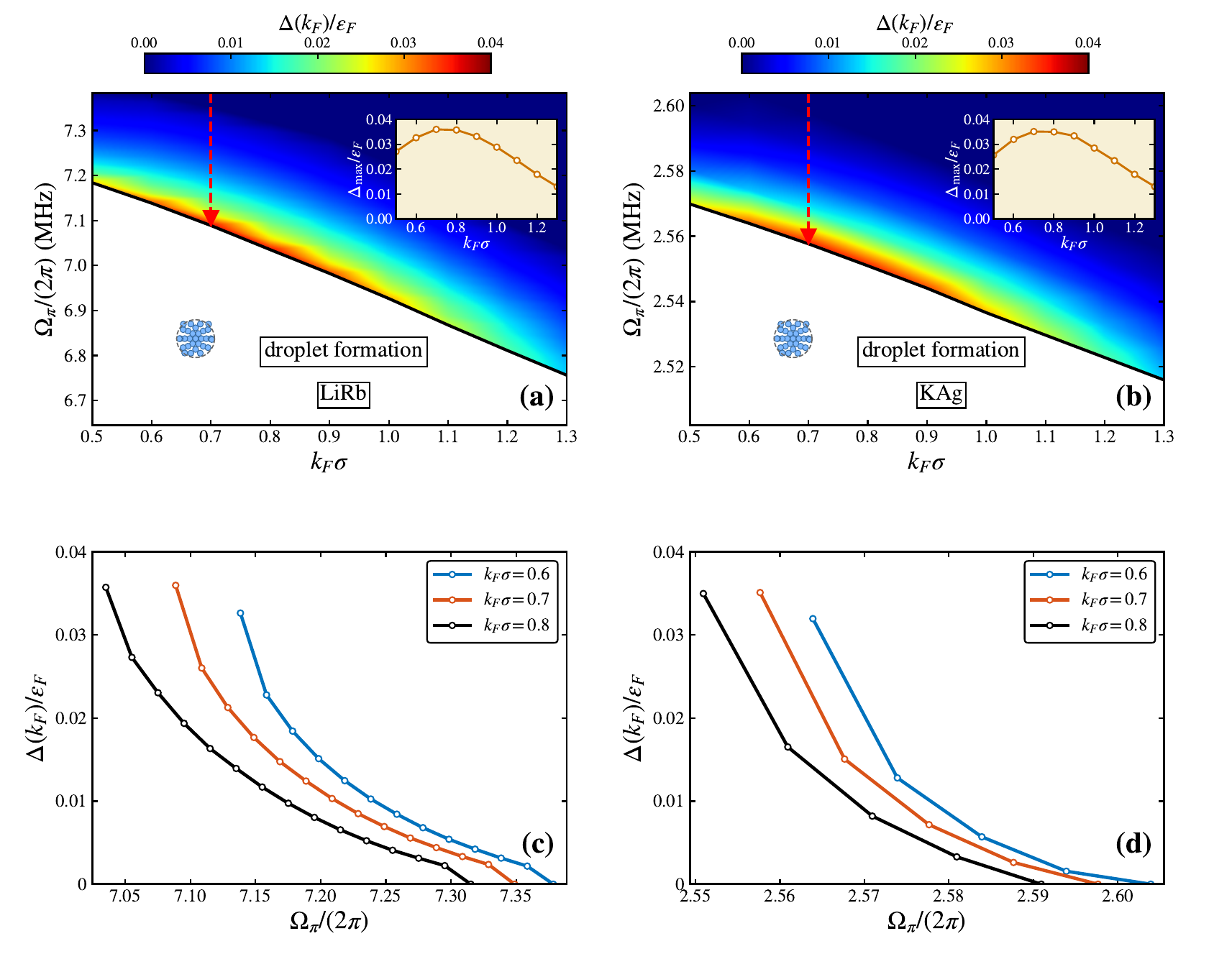}
    \caption{
Phase diagrams and $p$-wave superfluid gaps for LiRb and KAg.
(a)--(b) $p$-wave superfluid gap $\Delta(k_F)/\varepsilon_F$ in the
$k_F\sigma$--$\Omega_\pi/(2\pi)$ plane for LiRb and KAg, respectively,
calculated within FHNC using the quasi-2D interaction
$V_{\mathrm{2D}}$. The corresponding maximal gap as a function of
$k_F\sigma$ is shown in the inset.
(c)--(d) $p$-wave superfluid gap as a function of $\Omega_\pi/(2\pi)$ at
$k_F\sigma=0.6$, $0.7$, and $0.8$ for LiRb and KAg, respectively.
}
    \label{phasemol}
\end{figure}

\begin{figure}[t]
    \centering
    \includegraphics[width=\columnwidth]{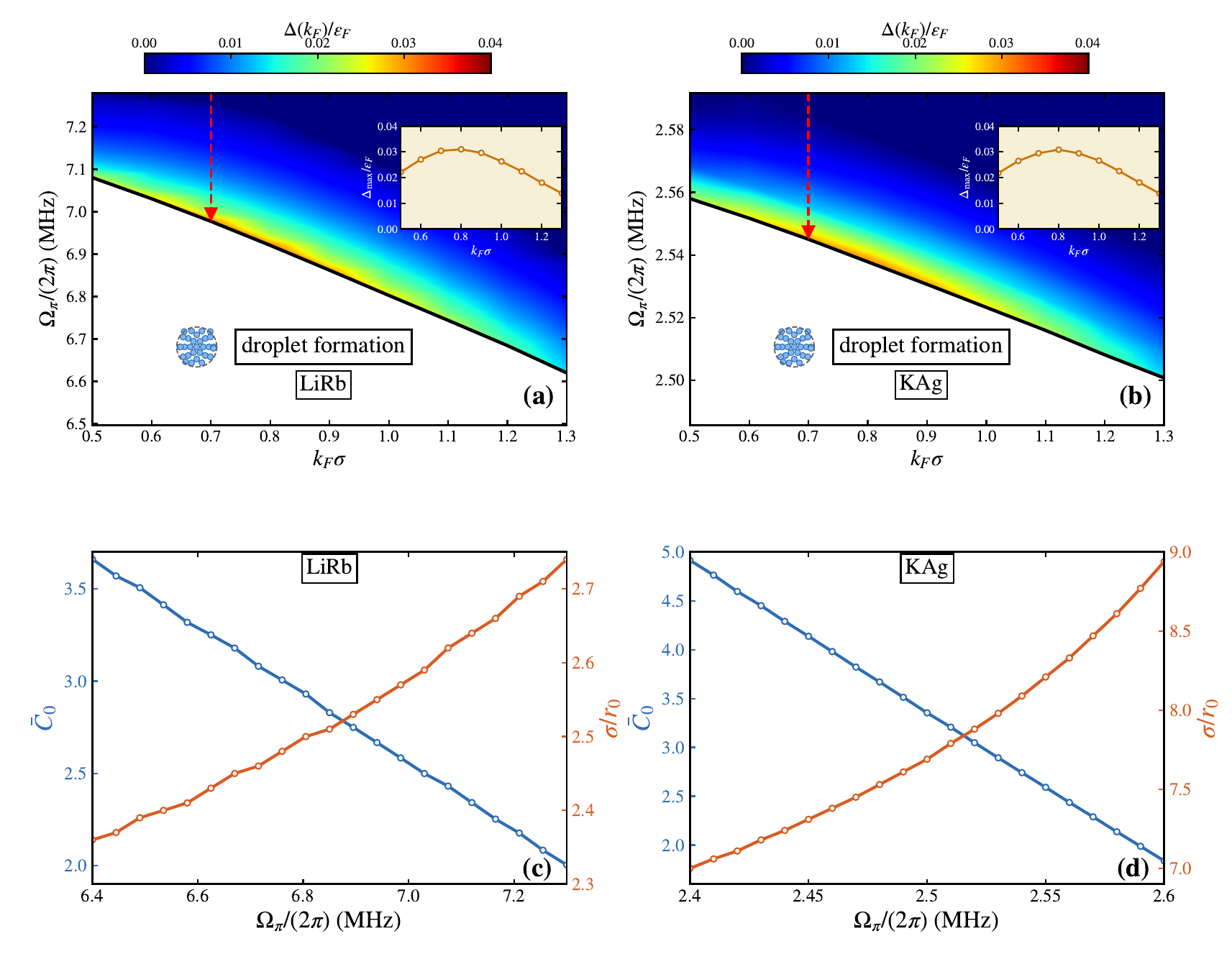}
    \caption{
Phase diagrams and $p$-wave superfluid gaps for LiRb and KAg, where the confinement lengths are $\sigma_z/r_0=1$ and $3$, respectively.
(a)--(b) $p$-wave superfluid gap $\Delta(k_F)/\varepsilon_F$ in the
$k_F\sigma$--$\Omega_\pi/(2\pi)$ plane for LiRb and KAg, respectively,
calculated within FHNC using the quasi-2D interaction
$V_{\mathrm{2D}}$. The corresponding maximal gap as a function of
$k_F\sigma$ is shown in the inset.
(c)--(d) Dimensionless interaction strength $\bar{C}_0$ and interaction length scale $\sigma/r_0$ as functions of the $\pi$-field Rabi frequency $\Omega_\pi/(2\pi)$ for LiRb and KAg, respectively.
}
    \label{phasemol2}
\end{figure}

Using the dual-microwave working points of Table~\ref{tab:dual_mw_parameters}, Figs.~\ref{phasemol}(a) and \ref{phasemol}(b) show the FHNC phase diagrams obtained directly from the quasi-2D interaction $V_{\mathrm{2D}}$, Eq.~\eqref{V2D}. The vertical axis is now the compensating-field Rabi frequency $\Omega_\pi$: decreasing $\Omega_\pi$ makes the net interaction more attractive, increases $\bar C_0$ [cf.~Fig.~\ref{gapC0sigma}(c) and~\ref{gapC0sigma}(d)], and eventually drives the system to the long-wavelength spinodal boundary associated with droplet formation. The stable and unstable regions therefore appear in the reverse vertical order from those in the $\bar C_0$ phase diagram of Fig.~\ref{phaseC3C6}(a).
The insets show the maximal superfluid gap at each $k_F\sigma$. For both molecules, the microscopic results closely follow the corresponding universal-model curve. The maximal gaps reach $\Delta_{\max}/\varepsilon_F\simeq0.036$ at $(\Omega_\pi/(2\pi),\bar C_0,k_F\sigma)=(7.09\,\mathrm{MHz},3.35,0.7)$ for LiRb and $\simeq0.035$ at $(2.56\,\mathrm{MHz},3.38,0.7)$ for KAg.

This agreement provides a nontrivial check of the universal reduction in the regime where pairing is strongest. As $\Omega_\pi$ is lowered, the microscopic $g_{\mathrm{dd}}$ and $S(k)$ exhibit the same characteristic evolution as in Fig.~\ref{phaseC3C6}: the correlation peak near the attractive minimum is progressively enhanced, while the low-$k$ slope of $S(k)$ increases and the sound velocity eventually vanishes at the phase boundary. The optimal pairing region is therefore set by proximity to the same correlated density instability in both the universal and microscopic descriptions.

Figures~\ref{phasemol}(c) and \ref{phasemol}(d) show the superfluid gap as a function of $\Omega_\pi/(2\pi)$ at $k_F\sigma=0.6$, $0.7$, and $0.8$ for LiRb and KAg, respectively. The comparison makes the experimental trade-off explicit: LiRb exhibits a broader microwave-control window but weaker collisional protection, whereas KAg has exceptionally small calculated loss but a stronger sensitivity of the effective interaction, and hence of the gap, to the compensating field. Together with the temperature and confinement scales in Table~\ref{tab:dual_mw_parameters}, these results favor an intermediate molecular regime when robustness is prioritized, while more strongly dipolar species become increasingly attractive as lower temperatures and improved microwave stabilization become available.

Finally, we examine to what extent the tight-confinement requirement can be relaxed, and the corresponding cost in pairing strength. We keep the remaining molecular and microwave parameters fixed and increase the confinement lengths to $\sigma_z/r_0=1$ for LiRb and $\sigma_z/r_0=3$ for KAg, substantially relaxing the condition $\sigma_z/r_{\min}\simeq0.2$ used above. As illustrated in Fig.~\ref{figpotential}(a), increasing $\sigma_z$ softens the short-range repulsive core of the quasi-2D interaction. The homogeneous fluid therefore reaches the long-wavelength spinodal instability at a weaker attraction, or equivalently at a smaller $\bar C_0$. This shift is evident from the phase diagrams in Figs.~\ref{phasemol2}(a) and \ref{phasemol2}(b), together with the dependence of $\bar C_0$ on $\Omega_\pi$ shown in Figs.~\ref{phasemol2}(c) and \ref{phasemol2}(d). As a consequence, the system can no longer approach the optimal pairing regime of the tightly confined case: the maximal gap is reduced to $\Delta_{\max}/\varepsilon_F\simeq0.031$ for both LiRb and KAg, as shown in the insets of Figs.~\ref{phasemol2}(a) and \ref{phasemol2}(b). The gain in experimental accessibility, however, is substantial. The required confinement frequencies are reduced to $\omega_z/(2\pi)\simeq39\,\mathrm{kHz}$ for LiRb and $2.7\,\mathrm{kHz}$ for KAg. Figure~\ref{phasemol2} therefore makes explicit the trade-off between confinement and pairing: substantially looser confinement can be achieved at the cost of a moderate reduction of the maximal superfluid gap.

\section{Conclusions and outlook}
\label{sec:conclusion}

We have identified the correlated many-body window in which a microwave-shielded molecular Fermi gas can support an appreciable $p$-wave gap while remaining stable against long-wavelength density fluctuations. A central simplification is provided by the universal quasi-2D interaction of Eq.~\eqref{eq:Veff}: under sufficiently tight confinement along the $z$ direction, the microscopic interaction is characterized by an interaction length $\sigma$ and a dimensionless coupling $\bar C_0$. FHNC-EL optimization then determines the correlation hole, density response, and stability boundary of the normal fluid, while correlated-basis theory describes pairing on top of this optimized correlated state. The strongest stable pairing occurs immediately on the homogeneous side of the spinodal boundary, around $k_F\sigma\simeq0.7$ and $\bar C_0\simeq3$--$4$, where the maximal gap reaches $\Delta_{\max}/\varepsilon_F\simeq0.036$. This location is the central many-body result: increasing the bare attraction does not monotonically improve pairing, but instead drives the system through the optimal pairing regime and eventually toward the spinodal instability associated with droplet formation.

Mapping this universal window back to microscopic dressing parameters turns the many-body result into a practical design criterion. With a single microwave field, the attractive tail and the repulsive shielding core originate from the same dressing mechanism, so reducing the attraction to the optimal many-body range simultaneously weakens collisional protection. Dual-microwave dressing removes this constraint: the $\sigma$ field can maintain a strong shielding barrier, while the $\pi$ field independently compensates the long-range attraction. Although a direct search over the four-dimensional space $(\Omega_\sigma,\delta_\sigma,\Omega_\pi,\delta_\pi)$ would be computationally demanding, the universal phase diagram strongly restricts the relevant parameter region. The microscopic search can instead be organized around four physical requirements---sufficient attraction for pairing, stability against droplet formation, suppression of collisional loss, and robustness against microwave fluctuations---with the requirement of experimentally feasible confinement along the $z$ direction providing an additional constraint. In this way, the FHNC-EL phase diagram and the constrained microscopic optimization together provide a systematic route for locating experimentally relevant working points rather than relying on a brute-force parameter search. An independent calculation based on neural-network quantum states, to be reported elsewhere, identifies superfluidity in the same parameter region and finds a self-bound droplet at stronger coupling beyond the FHNC spinodal boundary~\cite{leo2026}, providing a complementary check of the correlated many-body window identified here. 

The representative molecular species illustrate the resulting trade-offs. For NaK, simultaneously achieving sufficiently low loss and feasible confinement remains particularly demanding: the working point near the optimal pairing regime requires confinement on the megahertz scale. LiRb provides a more balanced regime. Under tight confinement optimized for the largest gap, $\omega_z/(2\pi)\simeq316\,\mathrm{kHz}$ and $\beta_{\mathrm{inel}}=4.8\times10^{-14}\,\mathrm{cm^3\,s^{-1}}$, while sub-percent control of the compensating field is required. Importantly, this confinement is not necessary if one accepts a moderate reduction of the pairing gap. Increasing the confinement length lowers the required frequency to about $39\,\mathrm{kHz}$, while the maximal gap decreases only from $\Delta_{\max}/\varepsilon_F\simeq0.036$ to $\simeq0.031$. KAg lies at the opposite limit: its large dipole moment yields exceptionally strong collisional protection, with $\beta_{\mathrm{inel}}=5.3\times10^{-16}\,\mathrm{cm^3\,s^{-1}}$, and the optimal working point requires only $\omega_z/(2\pi)\simeq24\,\mathrm{kHz}$. The confinement can be relaxed further to about $2.7\,\mathrm{kHz}$, again at the cost of reducing the maximal gap to approximately $0.031$. The principal limitation for KAg instead comes from the low temperature and increased microwave sensitivity. At $k_F\sigma\simeq0.7$, the optimal working point corresponds to $T_F\simeq4.6\,\mathrm{nK}$, placing a substantially stronger demand on cooling. For strongly dipolar molecules, this limitation can in principle be mitigated by increasing $\Omega_\sigma$: the resulting smaller interaction length $\sigma$ raises $T_F$ at fixed $k_F\sigma$ and, in the parameter regime considered here, further improves collisional shielding. The price is a larger compensating field and an increased sensitivity of the effective attraction to microwave-amplitude fluctuations. The optimization for strongly dipolar molecules therefore ultimately becomes a balance between achievable cooling and microwave stabilization. More generally, increasing the dipole moment enlarges the parameter space in which strong shielding and $p$-wave pairing can coexist, but the practical advantage is eventually limited by these two experimental capabilities. For reference, a $0.1\,\mathrm{dB}$ variation in microwave power corresponds to a Rabi-frequency variation of order $1\%$, so the sub-percent requirements identified here lie close to the scale at which active microwave stabilization becomes important.

The predicted zero-temperature pairing scale is comparatively large for a weak-pairing $p$-wave state. In superfluid $^3$He, $T_c/T_F$ is of order $10^{-3}$, corresponding to a zero-temperature gap $\Delta/\varepsilon_F$ of the same order, up to strong-coupling corrections~\cite{Vollhardt2013,Krotscheck2000}. The maximal value $\Delta_{\max}/\varepsilon_F\simeq0.036$ obtained here is therefore more than an order of magnitude larger than the characteristic scale in liquid $^3$He. This enhancement originates from the tunability of the microwave-dressed interaction: the attractive component can be brought close to the correlated stability boundary, while the shielding core suppresses short-range overlap. The present calculation is restricted to zero temperature. A weak-coupling conversion of the maximal gap suggests a pairing scale of order a few percent of $T_F$, but a quantitative determination of $T_c$ will require a finite-temperature treatment that incorporates the same strong correlations. Developing such a theory is a natural next step.

Taken together, these results establish a concrete design principle for molecular $p$-wave superfluidity: the interaction should not be made as attractive as possible, but instead tuned toward the correlated spinodal boundary from the stable side, while retaining a sufficiently large shielding core and adequate control of the microwave dressing fields. This criterion connects the microscopic control parameters of microwave-shielded molecules directly to the weak-pairing chiral $p$-wave regime and its associated non-Abelian excitations~\cite{ReadGreen2000,Ivanov2001,Nayak2008,CooperShlyapnikov2009}.

\section*{Acknowledgments}
We thank Fulin Deng, Leonard Bleiziffer, Lei Wang and Su Yi for useful discussions. This work was supported by the National Key Research and Development Program of China (Grant No. 2021YFA0718304) and the NSFC (Grants No. 12525413 and No. 12135018).

\begin{widetext}
\appendix

\section{Structure factors and exchange convolution}

\label{app:SFI}

For reference, the ideal-gas structure factors used in the FHNC calculation are
\begin{equation}
S_{F}(\mathbf{k})=\left\{ 
\begin{array}{c}
1-\frac{1}{2\pi }(4\arccos \frac{\bar{k}}{2}-\bar{k}\sqrt{4-\bar{k}^{2}}%
)\theta (2-\bar{k})\text{ \ \ (2D)} \\ 
1-\frac{1}{16}(\bar{k}+4)(\bar{k}-2)^{2}\theta (2-\bar{k})\text{ \ \ (3D)}%
\end{array}%
\right. ,
\end{equation}%
where $\bar{k}=k/k_F$. The exchange convolution $I(\mathbf{k},\mathbf{r})$ appearing in Eq.~\eqref{Xee} is
\begin{equation}
I(\mathbf{k},\mathbf{r})=\frac{\nu }{\rho }\int \frac{d^{d}p}{(2\pi )^{d}}%
e^{-i\mathbf{p}\cdot \mathbf{r}}\theta\!\left(k_F-|\mathbf k-\mathbf p|\right)
\theta\!\left(k_F-|\mathbf p|\right).
\end{equation}
and satisfies $I^{\ast }(\mathbf{k},\mathbf{r})=e^{i\mathbf{k}\cdot \mathbf{r}}I(\mathbf{k},%
\mathbf{r})$.

In two dimensions,
\begin{equation}
I(\mathbf{k},\mathbf{r})=\sum_{m\in \mathbb{Z}}i^{-m}e^{-im(\varphi -\varphi _{k})}c_{m}(%
\bar{k},\bar{r}),
\end{equation}%
where $\varphi _{k}$ and $\varphi $ denote the angles of $\mathbf{k}$ and $%
\mathbf{r}$, $\bar{r}=k_{F}r$, and%
\begin{eqnarray}
c_{m}(\bar{k},\bar{r}) &=&\frac{1}{\pi }\int_{0}^{1}\bar{p}d\bar{p}J_{m}(\bar{p}\bar{r}%
)\int_{0}^{2\pi }d\varphi _{p}e^{im\varphi _{p}}\theta (\cos \varphi _{p}-%
\frac{\bar{k}^{2}+\bar{p}^{2}-1}{2\bar{k}\bar{p}})  \notag \\
&=&\theta (1-\bar{k})[2\delta _{m0}\int_{0}^{1-\bar{k}}\bar{p}d\bar{p}J_{0}(\bar{p}%
\bar{r})+\frac{2}{\pi }\int_{1-\bar{k}}^{1}\bar{p}d\bar{p}J_{m}(\bar{p}\bar{r%
})\frac{\sin m\varphi _{L}}{m}]  \notag \\
&&+\theta (2-\bar{k})\theta (\bar{k}-1)\frac{2}{\pi }\int_{\bar{k}-1}^{1}%
\bar{p}d\bar{p}J_{m}(\bar{p}\bar{r})\frac{\sin m\varphi _{L}}{m}
\end{eqnarray}%
is determined by the angle $\varphi _{L}=\arccos [(\bar{k}^{2}+\bar{p}%
^{2}-1)/(2\bar{k}\bar{p})]$. For \(m=0\), the ratio is understood in the limiting sense,
\begin{equation}
\lim_{m\rightarrow 0}
\frac{\sin(m\varphi_L)}{m}
=
\varphi_L.
\end{equation}

In three dimensions,
\begin{equation}
I(\mathbf{k},\mathbf{r})=\sum_{l\geq 0}(2l+1)i^{-l}P_{l}(\cos \theta _{kr})c_{l}(%
\bar{k},\bar{r}),
\end{equation}%
where $\theta _{kr}$ is the angle between $\mathbf{k}$ and $\mathbf{r}$, and%
\begin{eqnarray}
c_{l}(\bar{k},\bar{r}) &=&\frac{3}{2}\int_{0}^{1}\bar{p}^{2}d\bar{p}j_{l}(%
\bar{p}\bar{r})\int_{-1}^{1}dxP_{l}(x)\theta (x-\frac{\bar{k}^{2}+\bar{p}%
^{2}-1}{2\bar{k}\bar{p}})  \notag \\
&=&\theta (1-\bar{k})[\frac{3}{2}\int_{0}^{1-\bar{k}}\bar{p}^{2}d\bar{p}%
j_{l}(\bar{p}\bar{r})\int_{-1}^{1}dxP_{l}(x)  \notag \\
&&+\frac{3}{2}\int_{1-\bar{k}}^{1}\bar{p}^{2}d\bar{p}j_{l}(\bar{p}\bar{r}%
)\int_{\frac{\bar{k}^{2}+\bar{p}^{2}-1}{2\bar{k}\bar{p}}}^{1}dxP_{l}(x)] 
\notag \\
&&+\theta (2-\bar{k})\theta (\bar{k}-1)\frac{3}{2}\int_{\bar{k}-1}^{1}\bar{p}%
^{2}d\bar{p}j_{l}(\bar{p}\bar{r})\int_{\frac{\bar{k}^{2}+\bar{p}^{2}-1}{2%
\bar{k}\bar{p}}}^{1}dxP_{l}(x).
\end{eqnarray}

\section{Quasi-two-dimensional interaction}
\label{2Dpotential}

We give the closed form of the quasi-two-dimensional interaction used in the numerical calculations. For molecules confined by a harmonic trap of size $\sigma_z$ along the $z$-direction, averaging the three-dimensional interaction over the density profile along the $z$-direction gives
\begin{eqnarray}
V_{\mathrm{2D}}(\rho)
&=&
\frac{\bar{C}_3}{\sqrt{2\pi}}
\int_{-\infty}^{+\infty}d\bar{z}\,
e^{-\bar{z}^{2}/2}
\frac{1}{\bar{r}^{3}}
\left(
2-3\frac{\bar{\rho}^{2}}{\bar{r}^{2}}
\right)
\notag\\
&&+
\frac{1}{\sqrt{2\pi}}
\bigg[
(c_{60}+c_{62}+c_{64})
\int_{-\infty}^{+\infty}d\bar{z}\,
\frac{e^{-\bar{z}^{2}/2}}{\bar{r}^{6}}
\notag\\
&&\qquad
-(c_{62}+2c_{64})\bar{\rho}^{2}
\int_{-\infty}^{+\infty}d\bar{z}\,
\frac{e^{-\bar{z}^{2}/2}}{\bar{r}^{8}}
+
c_{64}\bar{\rho}^{4}
\int_{-\infty}^{+\infty}d\bar{z}\,
\frac{e^{-\bar{z}^{2}/2}}{\bar{r}^{10}}
\bigg],
\end{eqnarray}
where
$\bar{C}_3=C_3/\sigma_z^3$,
$\bar{z}=z/\sigma_z$,
$\bar{\rho}=\rho/\sigma_z$, and
$\bar{r}=(\bar{\rho}^{2}+\bar{z}^{2})^{1/2}$.
The coefficients characterizing the short-range shielding interaction are
\begin{eqnarray}
c_{60}
&=&
\frac{1}{\sqrt{4\pi}}\bar{C}_{60}
-\frac{1}{4}\sqrt{\frac{5}{\pi}}\,\bar{C}_{62}
+\frac{9}{16\sqrt{\pi}}\,\bar{C}_{64},
\notag\\
c_{62}
&=&
\frac{3}{4}\sqrt{\frac{5}{\pi}}
\left(
\bar{C}_{62}
-\frac{3}{2}\sqrt{5}\,\bar{C}_{64}
\right),
\qquad
c_{64}
=
\frac{105}{16\sqrt{\pi}}\,\bar{C}_{64},
\end{eqnarray}
with $\bar{C}_{6l}=C_{6l}/\sigma_z^6$.

The contribution from the long-range dipolar interaction can be evaluated
analytically as
\begin{equation}
\begin{split}
&\int_{-\infty}^{+\infty}d\bar{z}\,
e^{-\bar{z}^{2}/2}
\frac{1}{\bar{r}^{3}}
\left(
2-3\frac{\bar{\rho}^{2}}{\bar{r}^{2}}
\right)
\\
&\qquad=
\frac{1}{2}e^{\bar{\rho}^{2}/4}
\left[
\bar{\rho}^{2}
K_{1}\left(\frac{\bar{\rho}^{2}}{4}\right)
-
(\bar{\rho}^{2}+2)
K_{0}\left(\frac{\bar{\rho}^{2}}{4}\right)
\right],
\end{split}
\end{equation}
where $K_m(x)$ is the modified Bessel function of the second kind. This
term determines the long-range attractive behavior of
$V_{\mathrm{2D}}(\rho)$.

The remaining three integrals associated with the shielding interaction
can also be evaluated analytically:
\begin{eqnarray}
\int_{-\infty}^{+\infty}d\bar{z}\,
\frac{e^{-\bar{z}^{2}/2}}{\bar{r}^{6}}
&=&
\frac{1}{8\bar{\rho}^{5}}
\bigg[
\sqrt{2\pi}\bar{\rho}(3-\bar{\rho}^{2})
\notag\\
&&\qquad
+\pi e^{\bar{\rho}^{2}/2}
(\bar{\rho}^{4}-2\bar{\rho}^{2}+3)
\operatorname{erfc}
\left(
\frac{\bar{\rho}}{\sqrt{2}}
\right)
\bigg],
\\
\int_{-\infty}^{+\infty}d\bar{z}\,
\frac{e^{-\bar{z}^{2}/2}}{\bar{r}^{8}}
&=&
\frac{1}{48\bar{\rho}^{7}}
\bigg[
\sqrt{2\pi}\bar{\rho}
(\bar{\rho}^{4}-4\bar{\rho}^{2}+15)
\notag\\
&&\qquad
-\pi e^{\bar{\rho}^{2}/2}
(\bar{\rho}^{6}-3\bar{\rho}^{4}
+9\bar{\rho}^{2}-15)
\operatorname{erfc}
\left(
\frac{\bar{\rho}}{\sqrt{2}}
\right)
\bigg],
\\
\int_{-\infty}^{+\infty}d\bar{z}\,
\frac{e^{-\bar{z}^{2}/2}}{\bar{r}^{10}}
&=&
\frac{1}{384\bar{\rho}^{9}}
\bigg[
\sqrt{2\pi}\bar{\rho}
(105-25\bar{\rho}^{2}
+5\bar{\rho}^{4}-\bar{\rho}^{6})
\notag\\
&&\qquad
+\pi e^{\bar{\rho}^{2}/2}
(\bar{\rho}^{8}-4\bar{\rho}^{6}
+18\bar{\rho}^{4}
-60\bar{\rho}^{2}+105)
\operatorname{erfc}
\left(
\frac{\bar{\rho}}{\sqrt{2}}
\right)
\bigg],
\end{eqnarray}
where $\operatorname{erfc}(x)$ denotes the complementary error function.

\end{widetext}

\bibliography{references}

\end{document}